\documentclass[letterpaper,aps,prl,twocolumn,floatfix,amsfonts,amssymb,notitlepage,superscriptaddress,longbibliography]{revtex4-2}

\pdfpageattr {/Group << /S /Transparency /I true /CS /DeviceRGB>>}

\usepackage{bbm}

\usepackage{amsmath}
\usepackage{amsthm}
\usepackage{amssymb}
\usepackage{graphicx}
\usepackage{tabularx}
\usepackage{color}
\usepackage{xcolor}
\usepackage[normalem]{ulem}
\usepackage[colorlinks=true, linkcolor=blue, citecolor=blue, urlcolor=blue]{hyperref}

\usepackage{placeins}

\newcommand{\be}{\begin{equation}}
\newcommand{\ee}{\end{equation}}
\newcommand{\bea}{\begin{eqnarray}}
\newcommand{\eea}{\end{eqnarray}}
\usepackage{bm}

\def\bs#1\es{\begin{split}#1\end{split}}	\def\bal#1\eal{\begin{align}#1\end{align}}

\usepackage{soul}
\usepackage[normalem]{ulem} % for \sout
\newcommand{\added}[1]{\textcolor{blue}{#1}}

\usepackage{braket}

\newcommand{\la}{\langle}
\newcommand{\ra}{\rangle}

\newcommand{\lp}{\left(}
\newcommand{\rp}{\right)}

\begin{document}

\title{Ballistic transport through 3D flat bands in rhombohedral graphene in an in-plane magnetic field
}

%\title{Rhombohedral graphene in an in-plane magnetic field: 3D flat bands and transport}

\author{Eran Sela}
\affiliation{Raymond and Beverly Sackler School of Physics and Astronomy, Tel Aviv University, Tel Aviv 69978, Israel}
\author{Rafi Bistritzer}
\affiliation{Raymond and Beverly Sackler School of Physics and Astronomy, Tel Aviv University, Tel Aviv 69978, Israel}

\author{Yuval Oreg}
	\affiliation{Department of Condensed Matter Physics, Weizmann Institute of Science, Rehovot, 76100, Israel}

%###################
%###################

\begin{abstract}
\noindent  
We explore how an in-plane magnetic field reshapes the transmitting channels of a finite slab of rhombohedral multilayer graphene and redistributes the current across its thickness. The two-terminal conductance counts these channels, while the voltage at each surface reflects the relative spectral weights of right- and left-moving states on the contacted layer. In sufficiently thick slabs or at sufficiently strong fields, the nearly flat bulk zeroth Landau level gives rise to a three-dimensional Hall effect with a plateau in the four-terminal Hall resistance. 
\end{abstract}
\maketitle

%###################
%###################
 \textbf{Introduction and main results:} Rhombohedral multilayer graphene (RMG) has emerged as a rich member of the van der Waals family, combining an unusual single-particle band structure with strong interaction effects. 
 In ABC-stacked graphene, the low-energy states are concentrated near the outermost layers, and their dispersion becomes flatter as the number of layers increases.
 This electronic structure provides a natural route to enhanced correlations and underlies the variety of symmetry-broken and topological phases observed in thin RMG~\cite{chen2019signatures,Shi_2020,Zhou_2021,Zhou_2021b,Han_2023,Han_2024,Han_2025,yang2025impact}.
 
From a three-dimensional perspective, RMG  provides a bulk realization of a nodal-line semimetal~\cite{Min_2008,Koshino_2009}. In a minimal model with interlayer hopping only between vertically aligned \(A\) and \(B\) sites in adjacent layers, the bulk bands touch along a Dirac spiral whose projection bounds a disk of radius $q_0$~\cite{Heikkil__2011}. 
See supplemental material (SM) for details. For each in-plane momentum $\mathbf q$, an $N$-layer slab maps onto an SSH chain with alternating hoppings. The chain is topological for $q<q_0$, and its end states form the weakly dispersing drumhead bands localized on the two surfaces. Remote hoppings modify the dispersion of these surface bands, which have been observed spectroscopically~\cite{pierucci2015evidence,Henck_2018,Zhang_2024}.  

An in-plane field $\mathbf {B} = B\hat {y} $ turns the nodal spiral into a family of two-dimensional Landau-level (LL) problems in the \(x\)-\(z\) plane, one for each conserved momentum \(q_y\).  For $|q_y|<q_0$, the minimal bulk model has a zeroth Landau level that remains at zero energy across most of this interval, whereas levels with $|q_y|>q_0$ disperse with $q_y$. The zeroth level thus lacks the free-particle dispersion along the field characteristic of a conventional three-dimensional metal. Its wavefunctions are localized along the stacking $\hat z$ direction; as their guiding centers approach the slab boundaries, the states become dispersive. The in-plane-field spectrum and the migration of low-energy states away from the surfaces were established for RMG by Tymczyszyn, Cross, and McCann~\cite{Tymczyszyn_2023}; related nearly flat levels have been studied in nodal-line semimetals~\cite{Rhim_2015,Molina_2018}.

\begin{figure*}[t]
\centering
\includegraphics[width=\textwidth]{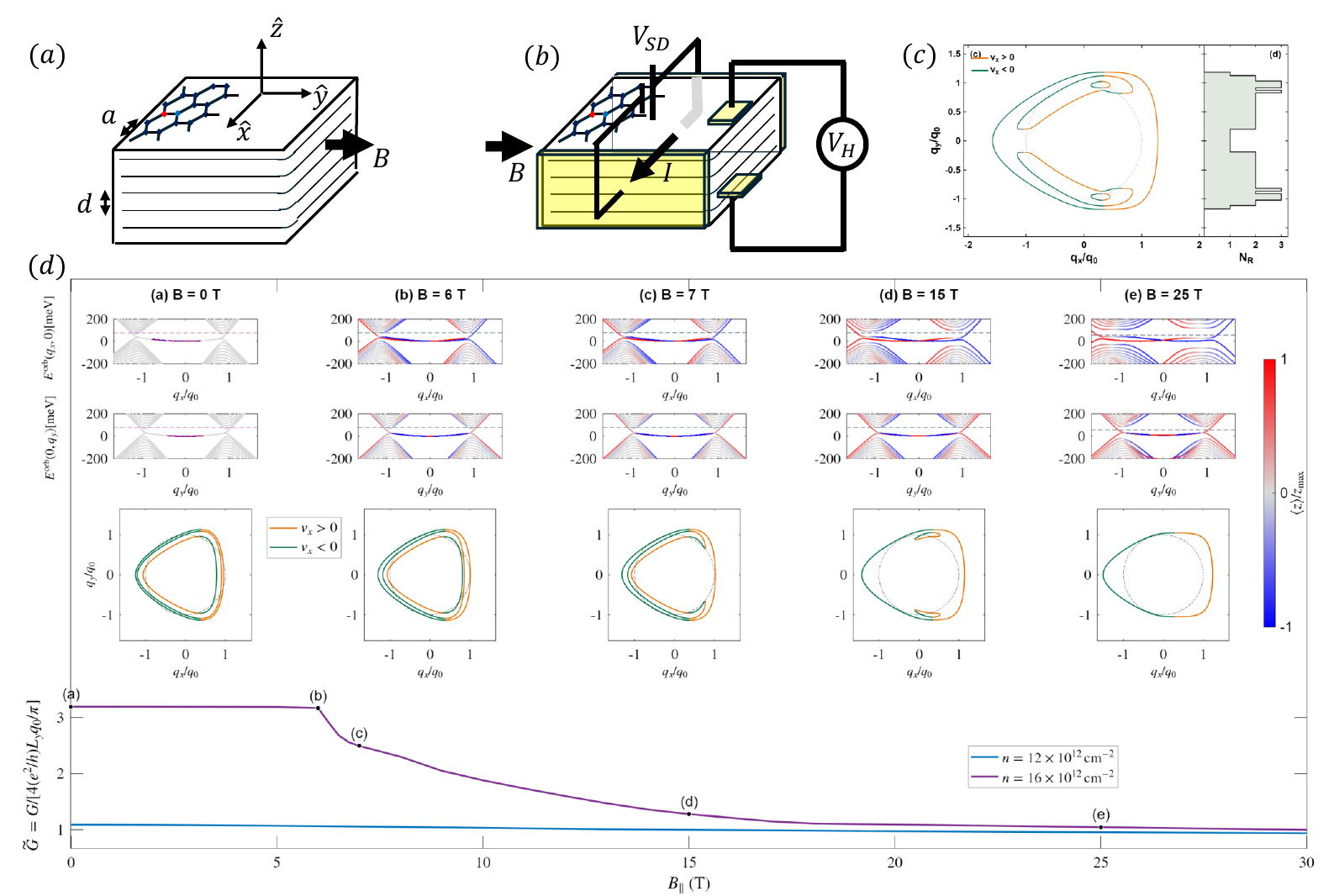}
\caption{ The main plot shows \(G=I/V_{\mathrm{SD}}\), normalized by \(4(e^2/h)q_0L_y/\pi\), for \(N=51\) at fixed carrier densities \(n=1.2\times10^{13}\,\mathrm{cm}^{-2}\) (blue) and \(n=1.6\times10^{13}\,\mathrm{cm}^{-2}\) (purple), including Zeeman splitting. The lower-density curve is nearly field independent, whereas the higher-density curve decreases above \(B\simeq6\,\mathrm{T}\) and approaches the same high-field plateau. The inset shows the device geometry, with \(\mathbf{B}=B\hat{\mathbf{y}}\) and a source–drain voltage \(V_{\mathrm{SD}}\) driving a current \(I\) along \(x\).
Panels (a)–(e) correspond to the marked fields on the higher-density curve. The top and middle rows show the orbital-dispersions \(E(q_x,0)\) and \(E(0,q_y)\), respectively. Color indicates the mean position \(\langle z\rangle/z_{\mathrm{max}}\), where \(z_{\mathrm{max}}=L_z/2\), and the horizontal dashed lines mark the chemical potential \(\mu(B)\). The bottom row shows the majority-spin Fermi contours, with orange and green segments denoting right-moving (\(v_x>0\)) and left-moving (\(v_x<0\)) states, respectively. The plot beside panel (c) gives the number of right-moving channels \(N_R(q_y)\) at \(B=7\,\mathrm{T}\). In panel (e), \(2q_c\) denotes the transverse width of the remaining Fermi contour.
} 
\label{fig:results}
\end{figure*}

In this manuscript, we study ballistic transport through RMG slabs in an in-plane magnetic field, calculating the two-terminal conductance, the layer-resolved current, and the voltage between weak voltage probes on opposite surfaces. Figure~\ref{fig:results} shows that the field dependence of the conductance varies strongly with carrier density. In some regimes the conductance is nearly field independent, while in others it decreases substantially before approaching a high-field plateau near
\begin{equation}
\label{eq:g0}
G_0 = 4 \frac{e^2}{h} \frac{q_0 L_y}{\pi}.
\end{equation}
Here \(L_y\) is the sample width along the field, and \(q_0L_y/\pi\) counts the transverse momenta in the interval \(|q_y|<q_0\). This scale corresponds to one right-moving channel per spin and valley for each such momentum.

In the regime we study, the current peaks near the center of the slab at weak fields. As the field increases, the central peak splits into two maxima that move toward opposite surfaces. The surface Hall voltage can vary nonmonotonically with magnetic field and approaches the two-terminal source--drain voltage only at high fields, see Fig.~\ref{fig:VH}. Thus, conductance, current distribution, and Hall voltage can evolve quite differently in the same slab.

We now show how the structure of the electronic states across the thickness of the slab shapes its transport response.

%%%%%%%%%%%%%%%%%%%

%%%%%%%%%%%%%%%%%%%%
 \textbf{Finite-slab model:}  We consider an $N$-layer rhombohedral graphene slab as in the inset of Fig.~\ref{fig:results} with in-plane magnetic field $\bm B=B\hat{\bm y}$, described by a vector potential $\bm A=Bz\hat{\bm x}$. We denote the layer coordinate $z_\ell=[(N+1)/2-\ell]d$ with $\ell =1,\dots, N$,  and $d=3.35 \AA$ the interlayer spacing. The drumhead radius is   $q_0=\gamma_1/(\hbar v)=(16.9\,\text{\AA})^{-1}$ where $\gamma_1$ is the leading interlayer hopping and $v=10^6$~m/s is the Dirac velocity.

The full finite-slab Hamiltonian for valley $\xi$ and spin $s$ is given by
\begin{equation}
H_{\xi s}(\bm q;B)
=
H_{\xi}^{(0)}(\bm q;B)
+
H_{\xi}^{(1)}(\bm q;B)
+
s\frac{g_s\mu_B|B|}{2}\mathbb I_{2N}.
\label{eq:model}
\end{equation}
Here $[H_{\xi}^{(0)}]_{\ell \ell'}=\delta_{\ell \ell'}v[\xi(\hbar q_x+eBz_\ell)\sigma_x+\hbar q_y \sigma_y] +(\gamma_1 \delta_{\ell \ell'-1}+h.c.)$, with Pauli matrices acting on the $A$ and $B$ sites of the graphene lattice, describes the minimal model. The second term $H_{\xi}^{(1)}$ accounts for remote hoppings; \(\gamma_2\) and \(\gamma_3\) modify the low-energy dispersion and Fermi-contour geometry, while \(\gamma_4\) introduces electron–hole asymmetry. The explicit form of the full  Hamiltonian is given in SM~\cite{supplement}. 
The last term in Eq.~\eqref{eq:model} describes the Zeeman splitting.

\begin{figure}[t]
\centering
\includegraphics[width=0.5\textwidth]{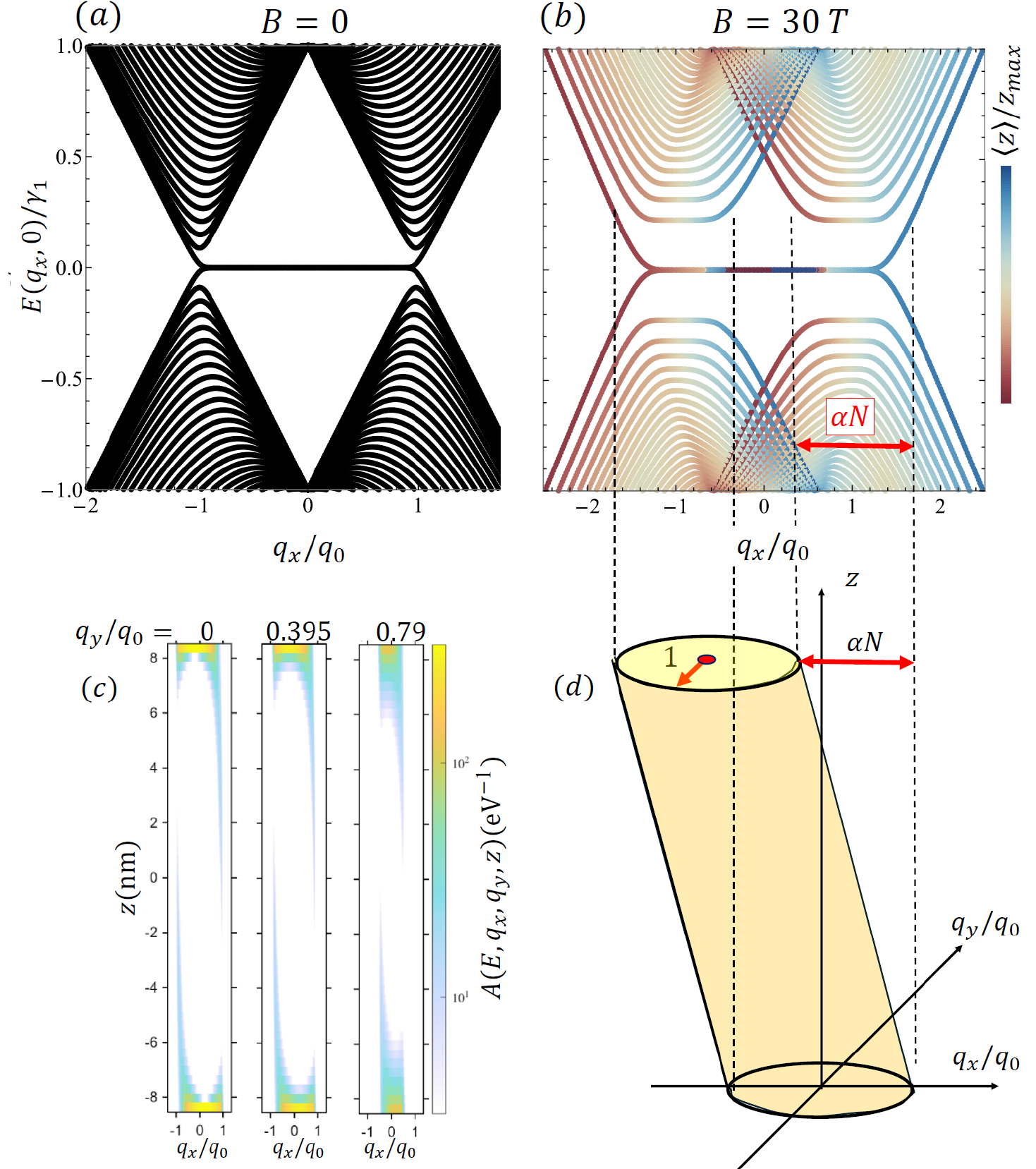}
\caption{ Band dispersion and spatial structure of electronic states in the minimal model for \(N=51\), retaining only the hopping amplitudes \(\gamma_0\) and \(\gamma_1\). (a,b) Dispersion versus \(q_x\) for \(q_y=0\), at \(B=0\) and \(B=30\,\mathrm{T}\), respectively. In (b), color indicates the mean position \(\langle z\rangle/z_{\max}\). (c) Layer-resolved spectral function \(A_\Gamma(E,q_x,q_y,z_\ell)\) at \(E=0\), for \(N=51\), \(B=10\,\mathrm{T}\), and valley \(\xi=1\), with Lorentzian broadening of 3meV (see Sec.~\ref{sec:zero_modes}). The three slices correspond to \(q_y/q_0=0,\ 0.395,\ 0.79\). The depletion of zero-energy spectral weight near the slab boundaries reflects hybridization between the bulk Landau states and the surface drumhead states. (d) Schematic tilted cylinder in \((q_x/q_0,q_y/q_0,z)\) space. Its circular bases represent the drumhead bands on opposite surfaces, whose centers are displaced relative to one another by approximately \(\alpha N\) along \(q_x/q_0\). Its lateral surface traces the guiding-center positions of the bulk zeroth Landau levels (see Sec.~\ref{eq:SM_3_SSH}). }
\label{fig:tilted}
\end{figure}

\begin{figure}[t]
\centering
\includegraphics[width=0.46\textwidth]{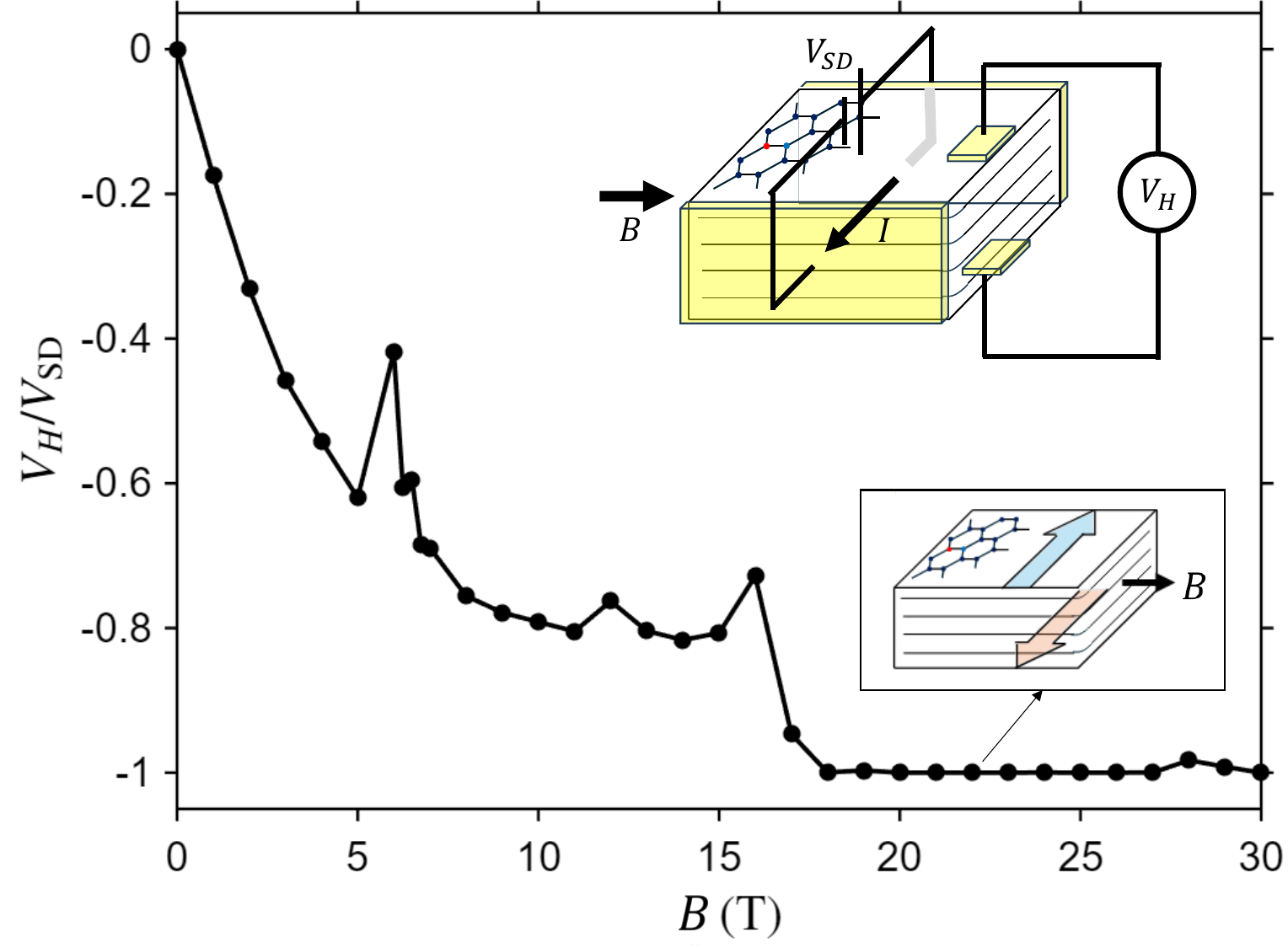}
\caption{The inset shows the setup with top and bottom voltage probes defining \(V_H\). The main plot shows \(V_H/V_{\mathrm{SD}}\) versus \(B\) for \(N=51\) and \(n=1.6\times10^{13}\,\mathrm{cm}^{-2}\), using the same parameters as the higher-density curve in Fig. 1. The ratio vanishes at \(B=0\), varies non-monotonically with field, and saturates at \(-1\) for \(B\gtrsim18\,\mathrm{T}\). In this regime the currents are peaked near the two surfaces (lower inset), see Sec.~\ref{se:layer_resolved_I}.
} 
\label{fig:VH}
\end{figure}

As evident from $H_\xi^{(0)}$, the field produces a layer-dependent momentum displacement of $d/\ell_B^2$ per layer. As displayed in Fig.~\ref{fig:tilted}, the total momentum shift of the drumhead in units of $q_0$ is   $\alpha N$, where
$\alpha=d/(\ell_B^2q_0)=0.92\times10^{-3} B[\mathrm{Tesla}]$
is a dimensionless measure of the magnetic field. 
For each fixed \(q_y\) within the projected nodal interval, the bulk low-energy theory describes Landau quantization in the \(xz\) plane. The LL-like bulk functions have energies \protect{$E_n = \pm \hbar/\ell_B  \sqrt{2 |n| v v_z(q_y)}$} where $v_z(q_y) = v d \sqrt{q_0^2 - q_y^2}$. Their spread along $z$ is $\delta_z (q_y) = \ell_B \sqrt{v_z(q_y)/v}$. It follows that $ \delta_z(0)  = \sqrt{d q_0} \ell_B$ where $ \sqrt{d q_0} \simeq 0.44$. These expressions break down near \(|q_y|=q_0\), where the two intersections of the nodal spiral with the plane of fixed \(q_y\) merge (Fig. S3). In a finite slab, the Landau levels acquire dispersion in \(q_x\) as their guiding centers approach either surface. Figure 2(a,b) compares the zero-field and finite-field spectra of the minimal model.

%%%%%%%%%%%%%%%%%%%%

In our system, level spacings arise from two sources: confinement along the stacking direction, $\Delta_N$, and Landau quantization, $\Delta_B$. Comparing them determines whether bulk-like Landau levels can form. At $B=0$ the level spacing at the nodal radius \(q=q_0\) defines the finite-size confinement scale $\Delta_N = \frac{\pi\gamma_1}{N}$, see Fig.~\ref{fig:tilted}(a). On the other hand, the first bulk LL energy at $q_y=0$  is given by $\Delta_B=\gamma_1\sqrt{2 \alpha}$, see Fig.~\ref{fig:tilted}(b). 
Bulk LLs therefore develop when $\Delta_B \gg \Delta_N$, i.e. $L_z \gg \frac{\pi}{2} \delta_z(0)$, namely when the wavefunctions  fit well within the bulk of the slab and remain weakly affected by its
boundaries.

 \textbf{Conductance:} We consider a two-terminal device of width $L_y$ (inset of Fig.~\ref{fig:results}), with current flowing along the $x$ direction between source and drain reservoirs held at chemical potentials $\mu\pm eV_{\rm SD}/2$. We calculate the current $I$ and the conductance $G=I/V_{\rm SD}$.

 In the ballistic regime, the Landauer--Imry--B\"uttiker scattering approach relates the conductance to the number of propagating channels at the Fermi energy. For a given transverse momentum $q_y$, each intersection of the dispersion with the Fermi energy at which the group velocity along $x$ is positive corresponds to a right-moving channel, contributing $e^2/h$ to the conductance under the assumption of perfect transmission. Denoting the number of such channels by $N_{R,\xi s}(E,q_y;B)$, we obtain, at zero temperature and in linear response,
\be
\label{eq:G}
\frac{G}{L_y}=\frac{e^2}{h}\sum_{\xi,s}\int\frac{dq_y}{2\pi}
N_{R,\xi s}\bigl(\mu(B),q_y;B\bigr),
\ee
where $\mu(B)$ is determined by the density  $n(\mu(B),B)$ (see Sec.~\ref{sec:DC_conductivity_LB}). An illustrative example of $N_R(q_y)$ is shown in Fig.~\ref{fig:results}(c).

The physical origin of this expression is the cancellation between the group velocity of each propagating state and its one-dimensional density of states. The conductance is therefore determined by the range of transverse momenta supporting propagating channels, weighted by their multiplicity. For a simple Fermi contour, this reduces to its projected width $2q_c$ along the $q_y$ direction [Fig.~\ref{fig:results}(e)]. More generally, a given \(q_y\) may support several right-moving channels, each of which contributes \(e^2/h\) to the conductance.

Consider the high-density conductance curve in Fig.~\ref{fig:results}. At weak fields, the nested Fermi contours give rise to several right-moving channels at the same \(q_y\). The total width of the contours projected onto the \(q_y\) axis, counted with channel multiplicity, changes little up to \(B\simeq 6\) T [Fig.~\ref{fig:results}(a,b)]. The conductance is therefore nearly constant even before bulk Landau levels form. Near \(6\) T, the inner contours reconnect, producing a kink in the conductance. As the resulting pockets shrink, the range of \(q_y\) supporting multiple right-moving channels narrows and the conductance decreases [Fig. 1(c,d)]. By \(25\) T, a single contour of transverse width \(2q_{c,s}\) remains for each spin \(s\), with one right-moving channel per valley at each \(q_y\) in that interval [Fig.~\ref{fig:results}(e)]. Zeeman splitting makes \(q_{c,s}\) spin dependent, but both widths are close to \(2q_0\), giving \(G\simeq G_0\). Throughout the field sweep, the contours are evaluated at the chemical potential \(\mu(B)\) required to keep the carrier density fixed, which can vary nonmonotonically (Fig.~\ref{fig: mu_B N51}).

\textbf{Hall voltage:} 
Eq.~(\ref{eq:G}) counts channels and is blind to where in the stack their current flows. The profile itself, obtained from the layer-resolved current operator (see Sec.~\ref{se:layer_resolved_I}), is governed by $L_z/\delta_z$. For $L_z\lesssim\delta_z$ the Fermi-level states are hybridized across the slab and the current is peaked at its center; for $L_z\gg\delta_z$ it splits into two maxima that move toward the surfaces, each penetrating a finite depth (Fig.~\ref{fig:LB_layer_current}). Surface spectral weight and surface current need not coincide, since a layer contributes according to the velocity operator rather than the probability weight. A direct measurement of current profile is experimentally challenging. A more accessible signature of the underlying 3D Hall effect is a four-terminal setup with voltage probes on the top and bottom surfaces (inset of Fig.~\ref{fig:VH}). However, as we now show, these probes measure neither the current profile nor the surface weight itself, but the directional selectivity of the surface spectral weight.

We assume weak coupling between the probes and the outermost layers of the slab as well as equal and incoherent coupling to the $p_z$ orbitals. Let $w_{pa}=\langle a|P_p|a\rangle$ be the weight of the Fermi-level eigenstate $a$ on the layer contacted by probe $p=\mathrm{t},\mathrm{b}$. Using the zero-current condition $I_p=0$, we then find that the layer resolved densities of right- and left-moving states (see Sec.~\ref{se:surface_voltage})
\begin{equation}
\nu_p^\pm = \sum_{\xi,s}\int\frac{dq_y}{(2\pi)^2}
\sum_{a:\,\pm v_{x,a}>0}
\frac{w_{pa}}{\hbar|v_{x,a}|},
\end{equation}
determine the floating-probe electrochemical potential 
$\mu_p = \lp \nu_p^+\mu_S+\nu_p^-\mu_D\rp/\lp \nu_p^++\nu_p^-\rp$. Here $a$ labels the Fermi crossings at fixed $q_y$, including the band index. The probe thus measures a weighted mixture of the source and drain electrochemical potentials. Defining the directional polarization of probe $p$, $\eta_p = \lp \nu_p^+-\nu_p^- \rp /\lp \nu_p^++\nu_p^- \rp$ the probe chemical potential is $\mu_p = \mu + \eta_p e V_{\rm SD}/2$.

The ratio of Hall voltage $eV_H=\mu_{\rm{t}}-\mu_{\rm{b}}$ and source drain voltage is therefore
\be
\frac{V_H}{V_{\rm SD}} = \frac{\eta_{\rm{t}}-\eta_{\rm{b}}}{2}
\label{eq:VH}
\ee
and the Hall resistance is $R_{\rm H} = V_H/I = \lp \eta_{\rm{t}}-\eta_{\rm{b}} \rp/(2G)$. 

At \(B=0\), time-reversal symmetry gives equal spectral weights to right- and left-moving states on each surface. Thus \(\eta_t=\eta_b=0\) and \(V_H=0\). At strong fields, right- and left-moving states are localized near opposite surfaces. When each probe samples only one propagation direction, one reads the source potential and the other reads the drain potential, giving \(\lvert V_H/V_{\rm SD}\rvert=1\) and \(\lvert R_H\rvert=1/G\).

At intermediate fields, each probe samples both propagation directions. Its voltage depends on which Fermi crossings are present at the fixed-density chemical potential \(\mu(B)\) and on the weight \(w_{pa}/\lvert v_{x,a}\rvert\) of each crossing on the contacted surface. Changes in the Fermi contours and in these weights can therefore produce a nonmonotonic field dependence of \(V_H\).

Figure~\ref{fig:VH} shows \(V_H/V_{\rm SD}\) for the high-density curve of Fig.~\ref{fig:results} (\(N=51\), \(n=1.6\times10^{13}\,\mathrm{cm}^{-2}\)). The upward feature near \(6\,\mathrm{T}\) coincides with the Lifshitz transition in Fig.~\ref{fig:results}. Further features appear near \(12\) and \(16\,\mathrm{T}\), before the ratio saturates at \(-1\) for \(B\gtrsim18\,\mathrm{T}\), approximately where \(G\) reaches its plateau near \(G_0\).

The End Matter examines the Hall conductivity \(\sigma_{xz}\), defined by \(j_x=\sigma_{xz}E_z\) for a uniform transverse field. Interestingly, it shows that the separation between counterpropagating modes continues to affect \(\sigma_{xz}\) even after the four-terminal Hall resistance has saturated.

 %%%%%%%%%%%%%%%%%%%%
%%%%%%%%%%%%%%%%%%%%

\textbf{Discussion:} We expect the single-particle framework used above to be adequate as long as $\mu$ lies above the zeroth Landau level, where the flat band is fully occupied and the Fermi-level states are dispersive. The Hartree potential generated by the surface-concentrated density acts as a layer-dependent potential of the displacement-field type (Sec.~\ref{eq:SM_1_T-B}). It shifts $\mu(B)$, $q_c$, the fields of the Lifshitz transitions and the current profile across the slab, but Eqs.~(\ref{eq:G})--(\ref{eq:VH}) apply unchanged to the self-consistent bands, and the high-field regime of one chiral pair per $q_y$ locked to opposite surfaces, with its plateau in $G$ and saturation of $V_H$, is a mean-field-stable feature. The same holds for an external displacement field, which tilts the Landau levels along $q_x$ (see Eq.~(\ref{eq:LL_D})) and leaves the picture intact as long as its screened drop across the Fermi-level states is smaller than $\Delta_B$.

Inversion symmetry implies that the two valleys contribute with the same sign to $G$, $V_H$ and $\sigma_{xz}$. A different behavior emerges for the pseudomagnetic field induced by a strain gradient along the stacking direction. For instance, strain applied to the bottom layer will propagate toward the top, and produce an in-plane pseudofield. A pseudofield preserves time reversal, however, and therefore has opposite sign in the two valleys. The conductance, which only counts channels, survives, but the chirality locking is reversed between $K$ and $K'$, so $V_H$ and $\sigma_{xz}$ vanish and the Hall response appears only in the valley channel, $V_H^{K}=-V_H^{K'}$. 

Once $\mu$ enters the zero-energy manifold, at lower density or higher field, the problem changes character: the degeneracy per volume is macroscopic and proportional to $B$ (Sec.~\ref{sec:zero_modes}). In the minimal model the zero-energy states lie on the surface of the tilted cylinder of Fig.~\ref{fig:tilted}(d). Its two bases are the drumhead bands of the two slab surfaces, displaced by $\pm\alpha Nq_0/2$ along $q_x$. Its wall is formed by the zeroth Landau levels, whose guiding center moves across the slab at the rate $d\langle z\rangle/dq_x=-\ell_B^2$. For $L_z\gg\delta_z$ this places zero-energy weight at every depth, and the ratio of surface to bulk zero-energy states is $\pi/2\alpha N$, decreasing with layer number and field.  The wall, however, is only partly covered; as the layer-resolved spectral function in Fig.~\ref{fig:tilted}(c) shows, a Landau branch leaving one drumhead hybridizes with the drumhead of the opposite surface before completing the traversal and moves away from zero energy.

We find that the number of states within each LL per unit volume is 
\be
\label{eq:rho_3D}
\rho_{3D}=\frac{q_0}{\pi^2 \ell_B^2}.
\ee
We include here a factor of 2 due to the positive and negative $q_x$ LL branches, see Sec.~\ref{sec:zero_modes}. This should be contrasted with the usual LL degeneracy per unit area  $\rho_{2D}=1/2 \pi \ell_B^2$. 
A fascinating question, which we leave for future work, is whether the correlated and topological phases found in thin rhombohedral graphene~\cite{chen2019signatures,Zhou_2021,Han_2023,Han_2024,Han_2025,yang2025impact} acquire a three-dimensional analogue in which the flat-band states are not confined to the surfaces.

%%%%%%%%%%%%%%%%%%%%
\begin{figure*}[t]
\centering
\includegraphics[width=0.9\textwidth]{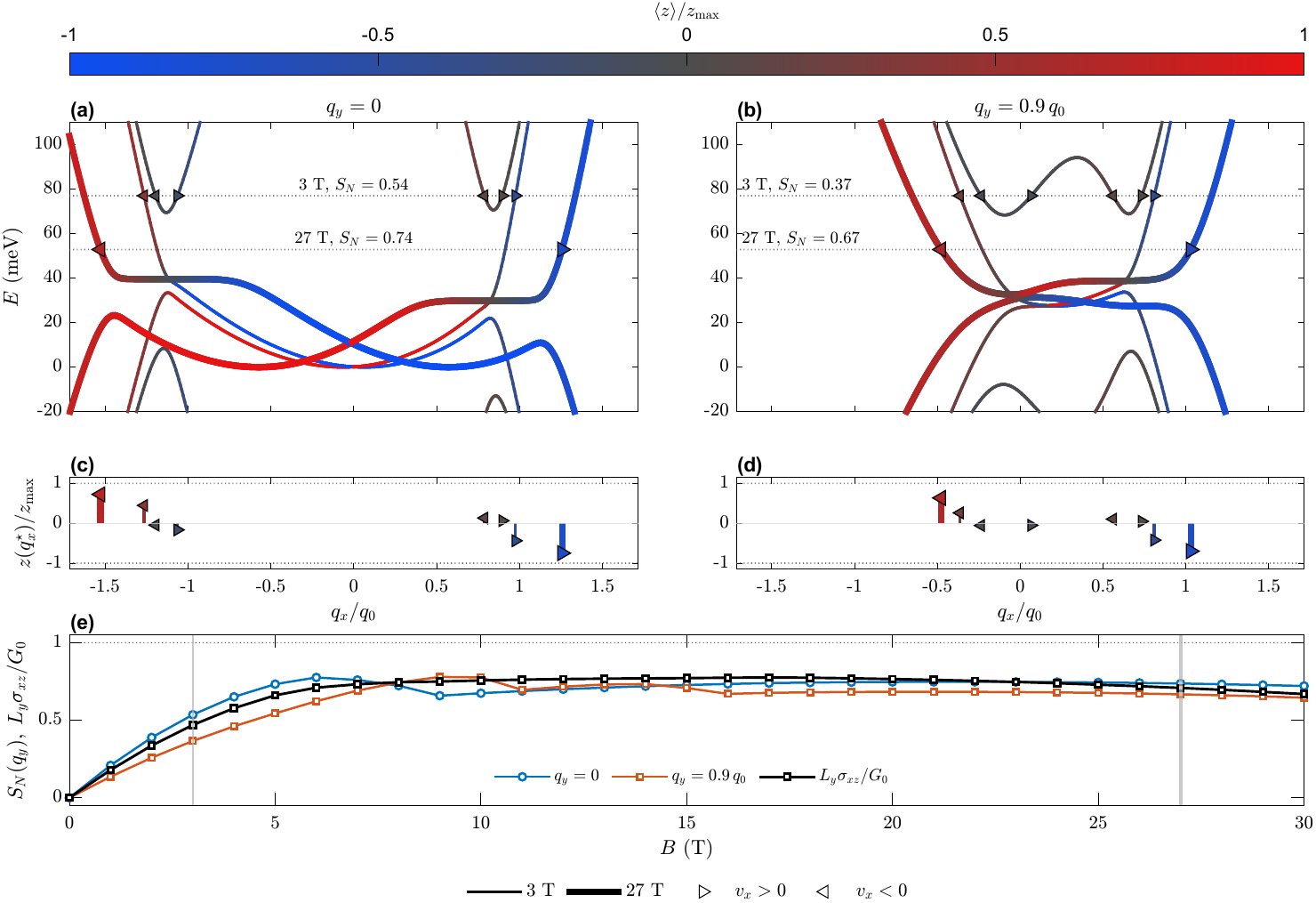}
\caption{(a,b) Band dispersions $E(q_x,q_y)$ in valley \(\xi=-1\) for \(N=51\), \(n=1.6\times10^{13}\,\mathrm{cm}^{-2}\), and \(q_y=0\) and \(0.9q_0\), respectively. Thin and thick lines correspond to \(B=3\,\mathrm{T}\) and \(27\,\mathrm{T}\). Color indicates the mean position \(\langle z\rangle/z_{\max}\). The dotted lines mark the majority-spin effective chemical potential. Triangles mark the corresponding Fermi crossings, point in the direction of propagation, and are colored according to \(z_a/z_{\max}\). The labels give \(S_N(q_y)\) at each field. (c,d) Positions \(z_a/z_{\max}\) of the Fermi crossings at momenta \(q_{x,a}\). At \(3\,\mathrm{T}\), the multiple crossings give a smaller \(S_N\) than the single counterpropagating pair at \(27\,\mathrm{T}\). (e) Field dependence of \(S_N(q_y)\) for the majority spin in valley \(\xi=-1\), at \(q_y=0\) and \(0.9q_0\), together with the normalized Hall conductance \(L_y\sigma_{xz}/G_0\), summed over both spins and valleys. Vertical lines mark \(3\) and \(27\,\mathrm{T}\). At high fields, \(L_y\sigma_{xz}/G_0\simeq0.7\) while \(V_H/V_{\mathrm{SD}}=-1\) (Fig.~\ref{fig:VH}). The bulk Hall conductivity remains sensitive to the separation between counterpropagating modes even after the surface probes become fully selective.} 
\label{fig:Kubo}
\end{figure*}

% %%%%%%%%%%%%%%%%%%%%

%%%%%%%%%%%%%%%%%%%%

\textbf{Acknowledgement:} ES acknowledges funding from European Research Council (ERC) under the European Union's Horizon 2020 research and innovation program under grant agreement No 951541 and from the Israel Science Foundation (ISF frant No. 1830/26).  RB acknowledges funding from the Israel Science Foundation (ISF Grant No. 2566/26). YO acknowledges funding from ISF
Grant No. 1914/24, ISF Quantum Science and Technology program (Grant No. 2074/19), the Minerva Stiftung and the Deutsche Forschungsgemeinschaft (DFG) via
CRC/Transregio 183.

%%%%%%%%%%%%%%%%%%%%

\section*{END MATTER} 

In the four-terminal setup of Fig.~\ref{fig:VH}, a current \(I_x\) is driven along the slab and the two surface probes measure a Hall voltage \(V_H\). Alternatively, one could calculate, using the standard linear-response Kubo formalism, the current response to a spatially uniform total transverse field \(E_z(\omega)\) and then take \(\omega\to0\). For a field amplitude \(|V_H|/L_z\), one might expect this procedure to reproduce the magnitude of the measured current \(I_x\).
However, for a finite slab, as we now explain, the two procedures correspond to different properties of the same states.

In the scattering approach leading to Eq.~\eqref{eq:VH}, each probe reads an electrochemical potential that depends on the scattering probability to the probe, which is set by the relative weights with which the probe couples to right- and left-moving states. As we now show, the response to a uniform field depends instead on where those states are centered across the slab. At fixed \(q_y\), when a single pair of states crosses the Fermi energy, its contribution is proportional to the separation of the two centers along \(z\). Thus the probes can already read the full source–drain voltage, \(|V_H|=|V_{\rm SD}|\), while that separation is still appreciably smaller than \(L_z\), see Fig.~\ref{fig:Kubo}. 
We note that in the thermodynamic limit, when $L_Z \rightarrow \infty$, the current-carrying modes are near the edge, and the quantized value of the Hall conductance for each $q_y$ is achieved, in agreement with the scattering approach.

For a prescribed uniform field \(E_z\), the Kubo formula gives the thickness-averaged current density \(\bar j_x=\sigma_{xz}E_z\), with (SM Sec.~\ref{sec:slab_conductivity})
\bea
\sigma_{xz} &=& -\frac{e^2}{\hbar L_z}\sum_{\xi,s}\int\frac{d^2q}{(2\pi)^2}
\sum_n f_n(\mathbf q)\,\frac{\partial z_n(\mathbf q)}{\partial q_x} \nonumber \\
&=& \frac{e^2}{h}\sum_{\xi,s}\int\frac{dq_y}{2\pi}\, S_N(q_y),
\label{eq:sigma_xz}
\eea
where $z_n=\la u_n|\hat Z|u_n\ra$, $S_N=-L_z^{-1}\sum_a{\rm sgn}(v_{x,a})\,z_a$, and $a$ runs over the Fermi crossings at fixed $q_y$, with positions $z_a$ and velocities $v_{x,a}$. The first equality also holds with remote hopping. The second is exact in the minimal model and remains accurate for the remote-hopping parameters considered here, for which the additional boundary term is numerically negligible (SM Sec.~\ref{sec:slab_conductivity}). 
Physically, the transverse field shifts each state energy by \(\delta E_n=eE_z z_n\), changing its velocity along \(x\) by \(\delta v_{x,n}=(eE_z/\hbar)\partial_{q_x}z_n\). Equation~\eqref{eq:sigma_xz} follows by summing the induced currents \(-e f_n\delta v_{x,n}\) and dividing by the slab thickness, with the occupations \(f_n\) held at their unperturbed values in the \(\omega\to0\) limit.
Similar expressions were obtained for the Hall conductivity of two-dimensional electron systems \cite{BartosRosenstein_1995, Streda_2007}.

Consider a fixed \(q_y\) with a single pair of Fermi crossings. For the field orientation shown in Fig.~\ref{fig:Kubo}, the left mover lies near the upper surface and the right mover near the lower surface. If their wavefunctions were centered exactly at the two surfaces, their contribution to Eq.~\eqref{eq:sigma_xz} would be
\[
S_N(q_y)=\frac{z_L-z_R}{L_z}=1.
\]
In the ideal limit, if one such pair is present for \(-q_c<q_y<q_c\) in each spin and valley sector, with no crossings outside this interval, Eq.~\eqref{eq:sigma_xz} gives
\be
\sigma_{xz}
=\frac{e^2}{h}\sum_{\xi,s}\int_{-q_c}^{q_c}\frac{dq_y}{2\pi}
=4\frac{e^2}{h} \frac{q_c}{\pi},                \label{eq:sigma_xz_ideal}
\ee
as expected for decoupled surfaces.
However, in a finite slab, the two crossing states are centered a distance of order \(\delta_z\) inside the surfaces. Their separation is therefore smaller than \(L_z\), reducing their contribution to the uniform-field Hall conductivity by a relative amount of order \(\delta_z/L_z\).

At fixed \(q_y\), the slab is a Hall strip in the \(xz\) plane. For \(|q_y|<q_0\) and \(\mu\) in the bulk gap above the zeroth Landau level, the two Dirac cones give a combined Chern number \(C(q_y)=1\) per spin and valley, fixing the net chirality at each surface. The two-terminal conductance \(G\) counts transmitting right movers with unit weight. The four-terminal Hall voltage in Eq.\eqref{eq:VH} depends on the relative contributions of right and left movers to the local density of states at each probe. When the two probes couple exclusively to modes arriving from opposite reservoirs, \(|V_H|=|V_{\rm SD}|\) and \(|R_H|=1/G\). Neither condition requires the mode centers to lie exactly at the surfaces. If each center lies a distance \(\delta_z\) inside its surface, Eq.~\eqref{eq:sigma_xz} gives \(S_N\simeq 1-2\delta_z/L_z\) even after the measured Hall voltage has saturated. In the thermodynamic limit, $\lim_{L_Z \rightarrow \infty}S_N = 1$, and as noted above, the Hall conductance for each $q_y$ is quantized.

Figure~\ref{fig:Kubo} shows the Fermi crossings and $S_N$ for the slab of
Fig.~\ref{fig:results} at $n=1.6\times10^{13}\,{\rm cm}^{-2}$. At 3~T, $\mu$ crosses several subbands near the center of the slab. At 27 T, the \(q_y=0\) slice shown in Fig.~\ref{fig:Kubo}(a) has a single pair of majority-spin crossings in valley \(\xi=-1\), with \(z_L\simeq+0.74z_{\max}\) and \(z_R\simeq-0.74z_{\max}\). Thus \(S_N\simeq0.74\), with each center about \(2.2\,\mathrm{nm}\) inside its surface.
At this field \(G\simeq G_0\) and \(V_H/V_{\rm SD}\simeq-1\), while \(L_y\sigma_{xz}/G_0\simeq0.7\) [Fig.~\ref{fig:Kubo}(e)].

%%%%%%%%%%%%%%%%%%%%

\bibliography{bibliography}

%%%%%%%%%%%%%%%%%%%%
%%%%%%%%%%%%%%%%%%%%

\clearpage
\onecolumngrid

% Supplemental page numbering: i, ii, iii, ...
\pagenumbering{roman}

% Supplemental section numbering: S.I, S.II, S.III, ...
\setcounter{secnumdepth}{3}
\setcounter{section}{0}
\renewcommand{\thesection}{S.\Roman{section}}

% Supplemental equation numbering: (S1), (S2), (S3), ...
\setcounter{equation}{0}
\renewcommand{\theequation}{S\arabic{equation}}

% Supplemental figure numbering: Fig. S1, Fig. S2, ...
\setcounter{figure}{0}
\renewcommand{\thefigure}{S\arabic{figure}}

% ============================================================
% Supplemental Material
% ============================================================

\begin{center}
{\large\bf Supplemental Material for}\\[0.5em]
{\large\bf ``Ballistic transport through 3D flat bands in rhombohedral graphene in an in-plane magnetic field
''}
\end{center}

\vspace{1em}

\begin{center}
Eran Sela, Rafi Bistritzer, and Yuval Oreg
\end{center}

%\vspace{0.5em}

\begin{center}
\small
\textit{Raymond and Beverly Sackler School of Physics and Astronomy,}
\textit{Tel Aviv University, Tel Aviv 69978, Israel}\\
\textit{Department of Condensed Matter Physics, Weizmann Institute of Science,}
\textit{Rehovot, 76100, Israel}
\end{center}

\vspace{1em}
\section{Band structure of RMG with in-plane magnetic field}
\label{eq:SM_1_T-B}
Here, we provide details of the model and band structure and include the effect of the displacement field. We follow the notation of Ref.~\cite{Tymczyszyn_2023}. 

The Hamiltonian depends on the in-plane momentum in each layer $\mathbf{k}_\parallel = (k_x, k_y)$ and is written as a $2N \times 2N$ matrix corresponding to $N$ layers and two atomic sites ($A$ and $B$) per layer. We modify the in plane terms $H_{AB}$ of the Hamiltonian by a Peierls substitution which introduces a phase
\be 
H_{AB}(\mathbf{k}_\parallel)=-\gamma_0 \sum_{j=1}^3 \exp \left( i \mathbf{k}_\parallel \cdot (\mathbf{R}_{B j}-\mathbf{R}_{A}) - \frac{i e}{\hbar} \int^{\mathbf{R}_{A}}_{\mathbf{R}_{B j}} \mathbf{A} \cdot \mathbf{d l} \right),
\ee 
where $\gamma_0$ is the tight binding parameter for intralayer hopping, $\mathbf{R}_{A}$ is the position of an $A$ atom, and $\mathbf{R}_{Bj}$ are the positions of the three neighboring $B$ atoms. 
The magnetic vector potential $\mathbf{A}=z (B_y,-B_x,0)$ is chosen to maintain in-plane translation symmetry, where $z$ is the direction perpendicular to the graphene layers. This gives 
\be 
H_{AB}(\mathbf{k}_\parallel)=-\gamma_0 \left[ e^{ \frac{i k_y a}{\sqrt{3}} - \frac{i e z a B_x}{ \sqrt{3} \hbar}} +2 e^{ \frac{-i k_y a}{2\sqrt{3}}+\frac{i e z a B_x}{2 \sqrt{3} \hbar}}  \cos\left( \frac{k_x a}{2} + \frac{ e z a B_y}{2 \hbar} \right) \right]. 
\ee 
Next we shift the in-plane wave vector as $\mathbf{q} = \mathbf{k}_\parallel - \mathbf{K}_\xi$ to describe deviations from one of the two valleys $\mathbf{K}_\xi= \xi(4 \pi/3a,0)$ where $\xi = \pm 1$ is the valley index. Then we expand the Hamiltonian for small $|\mathbf{q}|$ and for small $|\mathbf{B}|$ which is valid for $|\mathbf{q} a  | \ll 1$ and $e |z| a |\mathbf{B}| /\hbar \ll 1$. %The first condition holds at low energy, the second at low field. 
The resulting Hamiltonian is then
\be 
\label{eq:2prime}
H_{\xi}^{(0)}(\mathbf{q}) = \begin{pmatrix} 	0 &  v \pi_1^\dagger & 0 & 0 & \cdots & 0 & 0 \\ 	 v \pi_1 & 0 & \gamma_1 & 0 & \cdots & 0 & 0 \\ 	0 & \gamma_1 & 0 &  v \pi_2^\dagger & \cdots & 0 & 0 \\ 	0 & 0 &  v \pi_2 & 0 & \cdots & 0 & 0 \\ 	\vdots & \vdots & \vdots & \vdots & \ddots & \gamma_1 & 0 \\ 	0 & 0 & 0 & 0 & \gamma_1 & 0 &  v \pi_N^\dagger \\ 	0 & 0 & 0 & 0 & 0 &  v \pi_N & 0 \end{pmatrix} ,
\ee 
where $v=\sqrt{3} \gamma_0 a/2 \hbar$, $\pi_\ell (\mathbf{q}) = \xi (\hbar q_x+e  B_y z_\ell) + i (\hbar q_y - e  B_x z_\ell)$, 
and 
\begin{equation}
z_\ell = \left(\frac{N+1}{2}-\ell\right)d,
\qquad (\ell=1,\ldots,N),
\end{equation}
is the position of the $\ell$-th layer with $z=0$ at the middle layer. 
It is convenient to measure energies in units of $\gamma_1$ and wave vector in units of 
\be
q_0= \frac{\gamma_1}{v \hbar} = \frac{2}{\sqrt{3}a} \frac{\gamma_1}{\gamma_0} = (16.9 \AA)^{-1},  
\ee
yielding the dimensionless Hamiltonian 
\be 
\label{eq:H_dimless}
\frac{H_{\xi}^{(0)}(\tilde{\mathbf{q}})}{\gamma_1} = \begin{pmatrix} 	0 &  \tilde{\pi}_1^\dagger & 0 & 0 & \cdots & 0 & 0 \\ 	\tilde{\pi}_1 & 0 & 1 & 0 & \cdots & 0 & 0 \\ 	0 & 1& 0 & \tilde{\pi}_2^\dagger & \cdots & 0 & 0 \\ 	0 & 0 & \tilde{\pi}_2 & 0 & \cdots & 0 & 0 \\ 	\vdots & \vdots & \vdots & \vdots & \ddots & 1 & 0 \\ 	0 & 0 & 0 & 0 & 1& 0 & \tilde{\pi}_N^\dagger \\ 	0 & 0 & 0 & 0 & 0 & \tilde{\pi}_N & 0 \end{pmatrix} ,
\ee 
with
\begin{equation}
\widetilde{\pi}_\ell = \frac{v\pi_\ell}{\gamma_1}
= \xi\left[ \widetilde{q}_x
+ \left(\frac{N+1}{2}-\ell\right)\alpha_y
\right] + i\left[ \widetilde{q}_y
- \left(\frac{N+1}{2}-\ell\right)\alpha_x \right].
\end{equation}
Here $\tilde{q}_x=q_x/q_0$, $\tilde{q}_y=q_y/q_0$ and we defined the dimensionless magnetic field $\vec{\alpha}=\alpha \frac{\vec{B}}{|B|}$ with
\be
\alpha = \frac{d}{\ell_B^2 q_0}= 0.86 \cdot   10^{-3}  \times B[Tesla]. 
\ee
The various parameters that we  use are $v = 10^6$~m/s, $\gamma_1=0.39eV$, $\gamma_2=-0.017e V$, $\gamma_3=0.38eV$, $\gamma_4=0.14eV$, $a=2.46 \AA$ for the in-plane lattice constant, and $d=3.35 \AA$ for the interlayer spacing. %[With $a=2.46 \AA$ we have $\gamma_0=3.1eV$]. 
We next present few plots of the eigenvalues of the Hamiltonian including few extra terms.

\begin{figure*}[t]
\centering
\includegraphics[width=\textwidth]{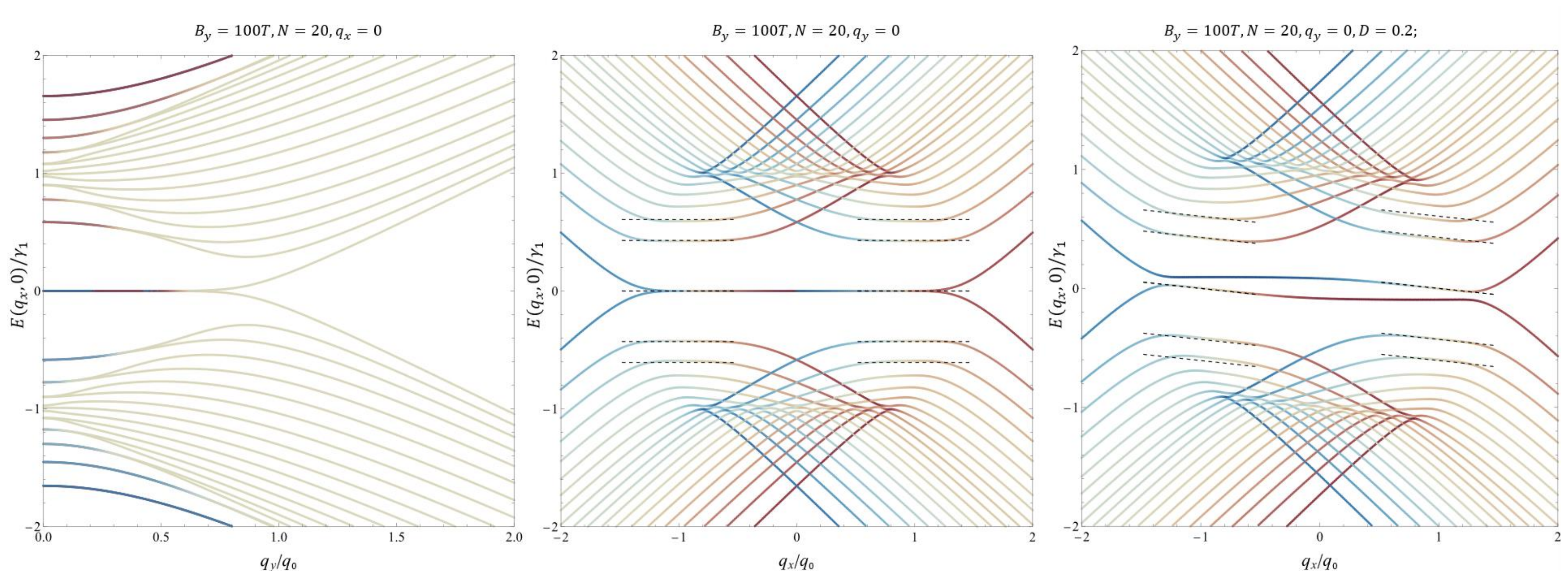}
\caption{Left panel: reproduction of the plot in Fig.~2b of Ref.~\cite{Tymczyszyn_2023} for the spectrum versus momentum along the field. Middle panel: spectrum versus momentum perpendicular to the field. Dashed lines are Landau levels, Eq.~(\ref{eq:LL_D}). Right panel: inclusion of a displacement field results in tilting of the Landau levels.
} 
\label{fig:1_SM}
\end{figure*}

The left panel of Fig.~\ref{fig:1_SM}  reproduces a plot in Fig.~2a in Ref.~\cite{Tymczyszyn_2023} for the spectrum versus momentum along the field, $q_y$. Plotting instead the spectrum versus momentum perpendicular to the field $q_x$ in the middle panel of Fig.~\ref{fig:1_SM}, shows flat LLs. The color scheme gives the expectation value of $z$. We can see 
a smooth change of color from the top layer (blue) to bottom layer (red). 

In order to see this layer dependence on the zero-th energy level, which is nearly doubly degenerate, it is convenient to add a  displacement field to the Hamiltonian $\frac{\delta 
H_{\xi}^{(0)}(\mathbf{q})_{\ell \ell'}}{\gamma_1}= D\delta_{\ell \ell'} \frac{ \frac{N+1}{2}-\ell}{N-1} \sigma^0$, %\added{(corresponds to N+1 layers)} 
such that $D$ is the potential difference between the boundary layers in units of $\gamma_1$. Using the results derived in the next sections, see Eq.~(\ref{eq:spec}) and Eq.~(\ref{eq:z_0}) for the average $z-$coordinate versus $q_x,q_y$, the resulting spectrum is  
\be
\label{eq:LL_D}
E_{n\tau} (q_x,q_y) \simeq
\operatorname{sgn}(n)\sqrt{2\alpha|n|}
\left[1-\left(\frac{q_y}{q_0}\right)^2\right]^{1/4}
-\frac{D}{(N-1)\alpha}
\left[
\frac{q_x}{q_0}
-\tau\sqrt{1-\left(\frac{q_y}{q_0}\right)^2}
\right].
\ee
%\be
%\label{eq:LL_D}
%\epsilon_n (q_x,q_y)={\rm{sgn}}(n) (1-(q_y/q_0)^2)^{1/4} \sqrt{2 \alpha |n|} + \frac{D}{N} \frac{1}{\alpha}[(q_x/q_0) +\tau],~~~(\tau=\pm 1).
%\ee
We can clearly see a tilting of the LLs in the right panel of Fig.~\ref{fig:1_SM} for $n \ne 0$. For $n=0$ we can see that the zero energy state, as function of $q_x$, consists of few segments: two segments are zero-th LLs, which tilt with $D$, and other segments which do not tilt but rather change by a constant $\pm D/2$, and correspond to the drumhead surface states. The dashed lines Fig.~\ref{fig:1_SM}  in the middle and right panels correspond to Eq.~(\ref{eq:LL_D}). The horizontal extent of each LL is taken from the relation $\frac{d z}{d q_x}=-\ell_B^2$  to be $-\Delta q_x/2<\lp q_x -\tau\sqrt{q_0^2-q_y^2} \rp <\Delta q_x/2$ where $\Delta q_x/ q_0 = (N-1) \alpha$.

%%%%%%%%%%%%%%%%%%%%

\begin{figure*}[t]
\centering
\includegraphics[width=.6\textwidth]{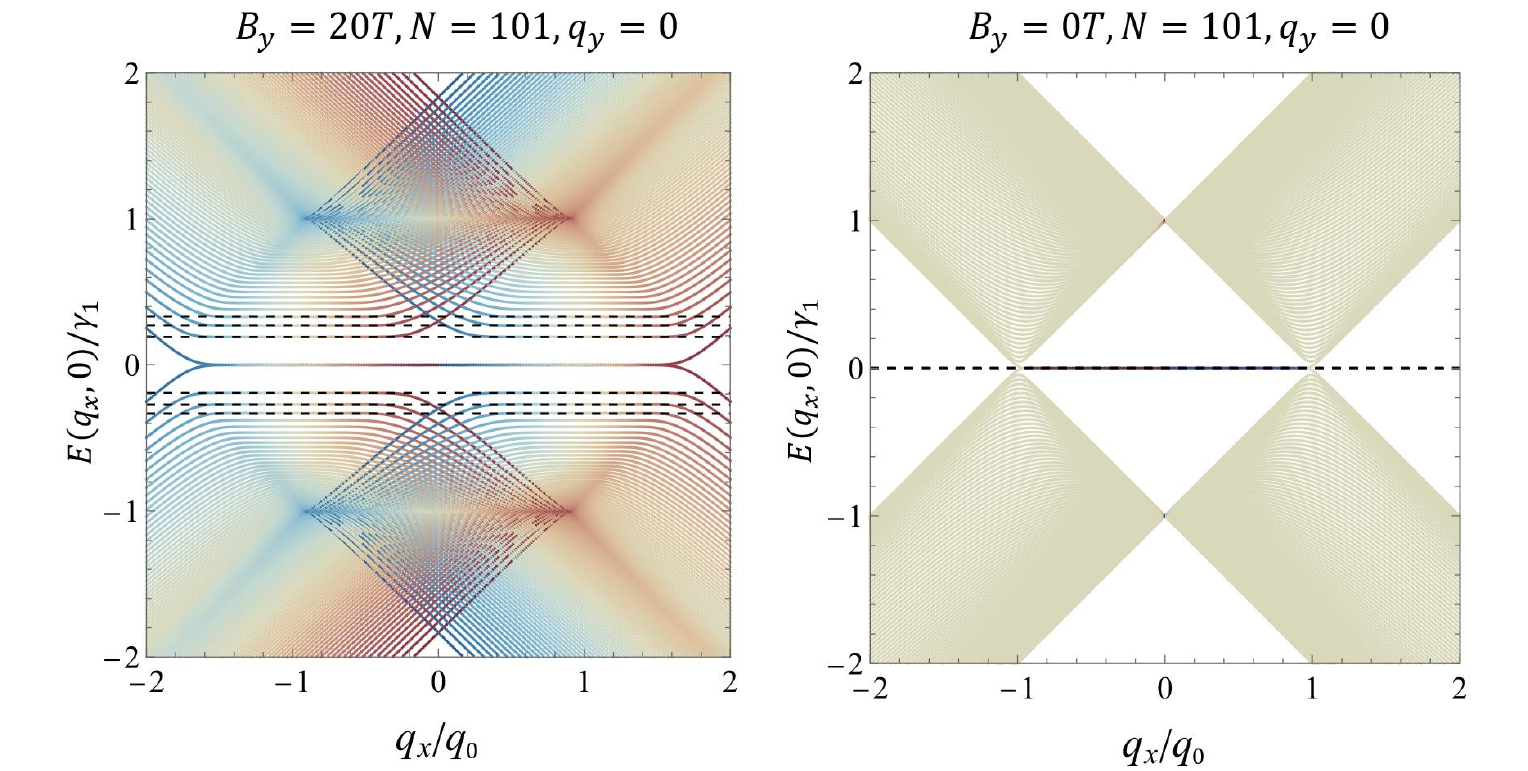}
\caption{Left panel: spectrum versus momentum perpendicular to the field showing clearly developed Landau levels for $N=101$ and $10T$ \added{(20T)} where the zero field level spacing $\Delta_N$ is much smaller than the Landau level gap $\Delta_B$. Right panel: the zero field drumhead dispersion.
} 
\label{fig:2_SM}
\end{figure*}

%%%%%%%%%%%%%%%%%%%%
These Landau levels strongly develop as the number of layers increases, see Fig.~\ref{fig:2_SM} for an extreme limit in which $\Delta_B \gg \Delta_N$.
%%%%%%%%%%%%%%%%%%%%

%%%%%%%%%%%%%%%%%%%%

\begin{figure*}[t]
\centering
\includegraphics[width=.8 \textwidth]{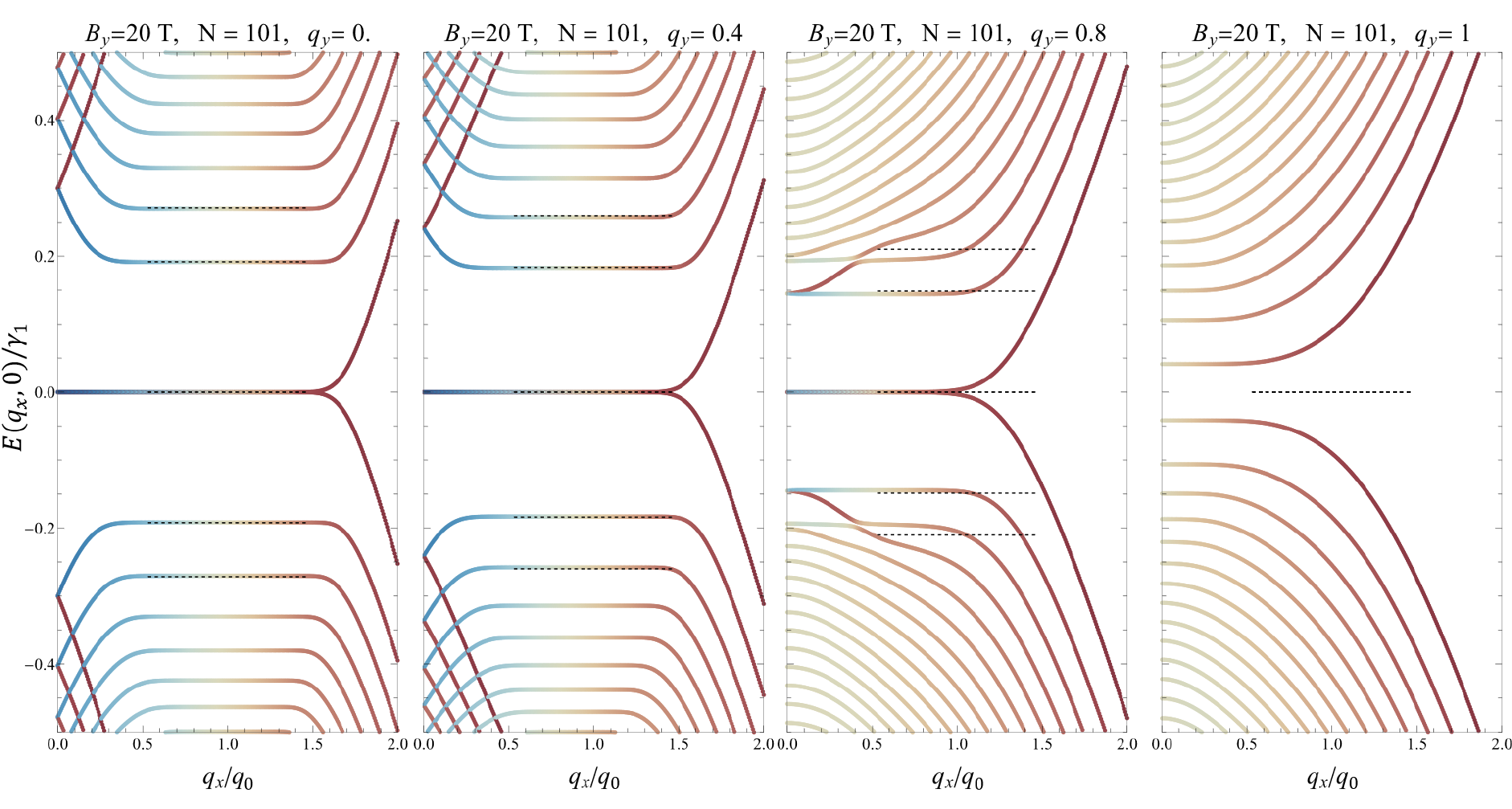}
\caption{Spectrum restricted to the first few Landau levels displaying the closure of the gap upon increasing $q_y$ according to Eq.~(\ref{eq:LL_D}).
Low-energy spectrum of the minimal model versus \(q_x\) for \(N=101\), \(B_y=20\,\mathrm{T}\), and increasing \(q_y\). Dashed lines show the bulk Landau-level energies from Eq. (\ref{eq:LL_D}) at \(D=0\). The level spacing decreases as \(q_y\) approaches \(q_0\). Near \(q_y=q_0\), the two nodal intersections merge and the separate-cone approximation breaks down; the finite-field spectrum retains a nonzero gap.
} 
\label{fig:3_SM}
\end{figure*}

Finally, the closing of the Landau level gap upon bringing the momentum parallel to the field towards the drumhead radius $q_0$ is shown in Fig.~\ref{fig:3_SM}.

%%%%%%%%%%%%%%%%%%%%

{\bf{Inclusion of long range hopping parameters:}} we ask whether  the LLs persist when considering the next-longer range hopping parameters in the Hamiltonian. The full Hamiltonian including these terms ($\gamma_2,\gamma_3,\gamma_4$) is
\be 
\label{eq:blockdiagonal}
H_{\xi}^{(0)}(\mathbf{q})+H_{\xi}^{(1)}(\mathbf{q}) = \begin{pmatrix} 	  \mathcal{A}_1 & \mathcal{B}_{3/2} & \mathcal{C} & \cdots & 0 & 0 \\ 	 \mathcal{B}^\dagger_{3/2} & \mathcal{A}_2  & \mathcal{B}_{5/2} & \cdots & 0 & 0 \\ 	 \mathcal{C}^\dagger & \mathcal{B}^\dagger_{5/2} & \mathcal{A}_3  & \cdots & 0 & 0 \\ 	\vdots & \vdots & \vdots & \ddots & \mathcal{B}_{N-3/2}  & \mathcal{C} \\ 	 0 & 0 & 0 & \cdots & \mathcal{A}_{N-1} & \mathcal{B}_{N-1/2} \\ 	 0 & 0 & 0 & \cdots & \mathcal{B}^\dagger_{N-1/2} & \mathcal{A}_{N} \end{pmatrix}, 
\ee 
where each element represents a 2*2 matrix,
\bea
\mathcal{A}_\ell=\begin{pmatrix} 	 0 & v \pi_\ell^\dagger  \\ 	  v \pi_\ell & 0 \end{pmatrix} ;~~~\mathcal{B}_\ell=\begin{pmatrix} 	 - v_4 \pi_\ell^\dagger &  v_3 \pi_\ell  \\ 	 \gamma_1 & - v_4 \pi_\ell^\dagger \end{pmatrix}  ;~~~  
\mathcal{C}=\begin{pmatrix} 	 0 & \frac{\gamma_2}{2}  \\ 	 0 & 0 \end{pmatrix} ;~~~~ v_i= \frac{\sqrt{3}a \gamma_i}{2\hbar}.
\eea
Rewriting this Hamiltonian in units of $\gamma_1$ and versus dimensionless momentum as in Eq.~(\ref{eq:H_dimless}), we obtain $\frac{H_{\xi}^{(0)}(\tilde{q})+H_{\xi}^{(1)}(\tilde{q})}{\gamma_1} $ having a block diagonal form exactly as in Eq.~(\ref{eq:blockdiagonal}) with $(\mathcal{A}_\ell,\mathcal{B}_\ell,\mathcal{C}_\ell) \to $
\bea
\tilde{\mathcal{A}}_\ell=\begin{pmatrix} 	 0 &  \tilde{\pi}_\ell^\dagger  \\ 	  \tilde{\pi}_\ell & 0 \end{pmatrix} ;~~~\tilde{\mathcal{B}}_\ell=\begin{pmatrix} 	 - \frac{\gamma_4}{\gamma_0}\tilde{\pi}_\ell^\dagger &  \frac{\gamma_3}{\gamma_0}\tilde{\pi}_\ell  \\ 	 1 & - \frac{\gamma_4}{\gamma_0}\tilde{\pi}_\ell^\dagger \end{pmatrix}  ;~~~
\tilde{\mathcal{C}}=\begin{pmatrix} 	 0 & \frac{\gamma_2}{2 \gamma_1}  \\ 	 0 & 0 \end{pmatrix}.
\eea

\begin{figure}
\centering
\includegraphics[width=1\textwidth]{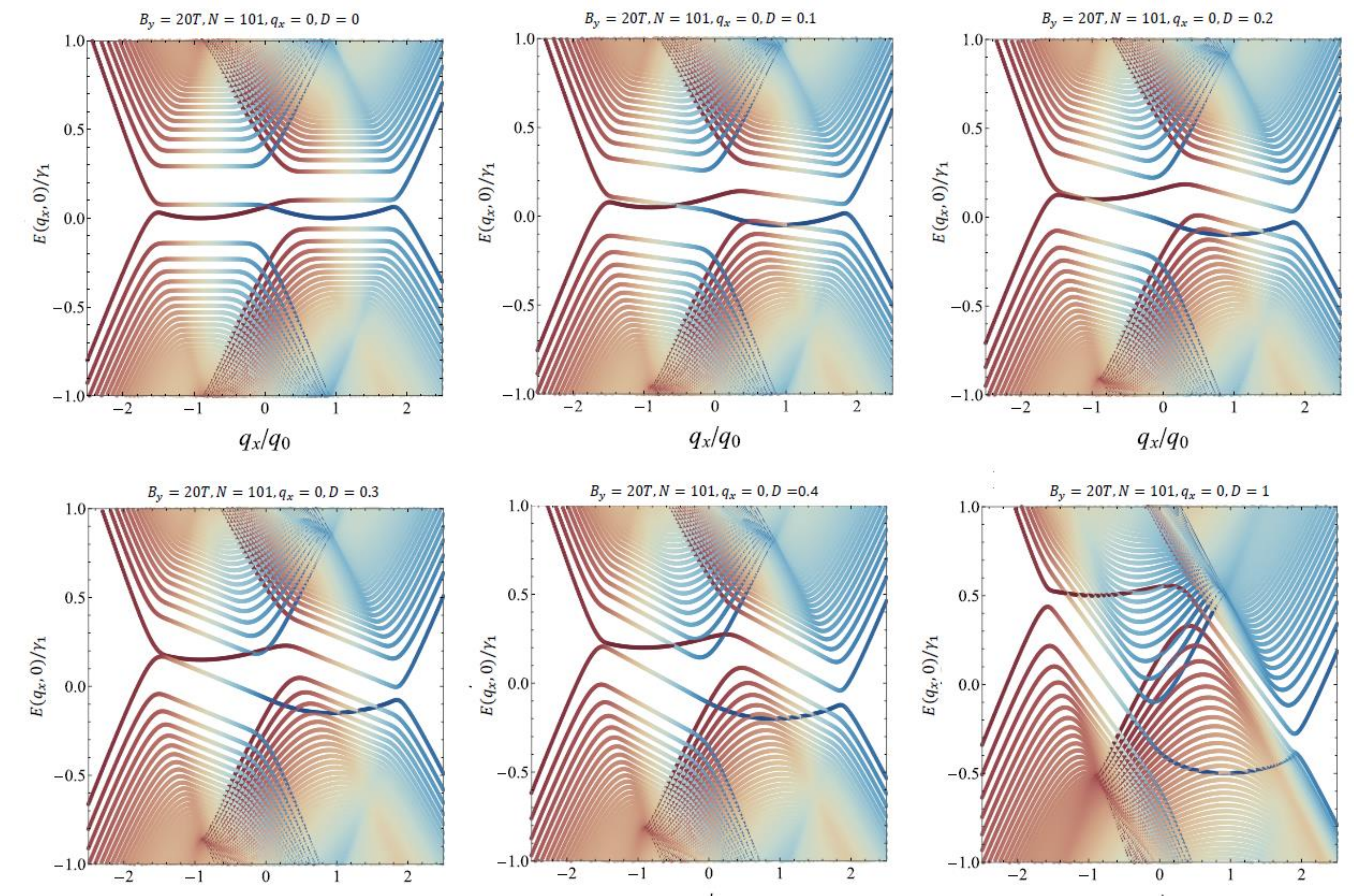}
\caption{Spectrum including $\gamma_2, \gamma_3$ and $\gamma_4$. At $D=0$ the 3D Landau levels retain nearly flat bulk-like portions while their energies shift due to the extra hopping parameters. Upon increasing $D$ the LLs tilt. Note a level crossing between $D=0.2$ and $D=0.3$. 
} 
\label{fig:3_gamma345}
\end{figure}

\FloatBarrier

\section{Semiclassical solution}
\label{eq:SM_2_semiclassical}
Here, we derive the 3D LLs from a semiclassical solution of the low-energy Dirac spiral Hamiltonian.

\begin{figure}
\centering
\includegraphics[width=.5\textwidth]{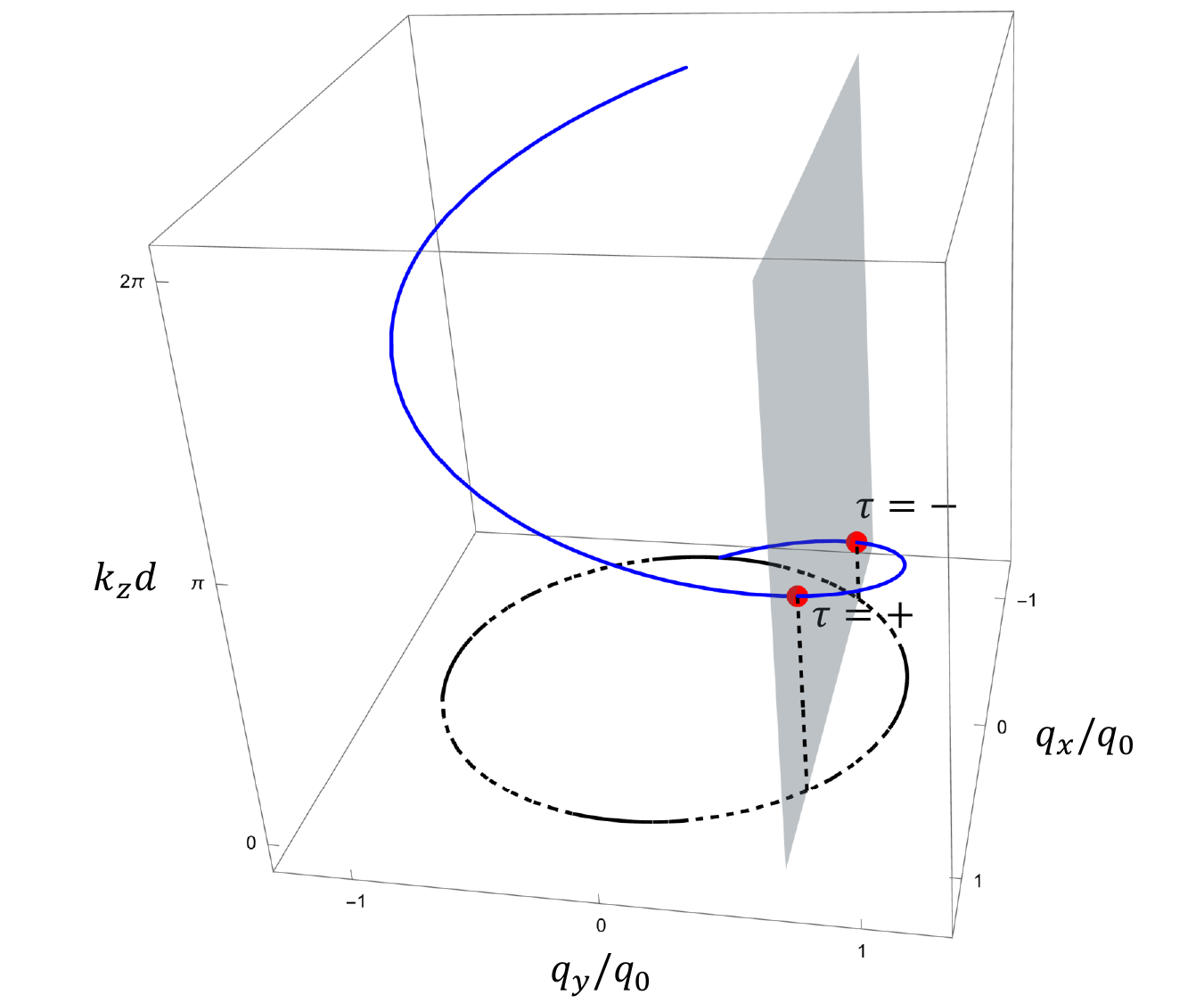}
\caption{Dirac spiral given by Eq.~(\ref{eq:nodal}) and its intersection with a plane of fixed $q_y=0.7 q_0$. The two points are denoted by $\tau=\pm 1$ according to the sign of $q_x$.
} 
\label{fig:helix}
\end{figure}

Imposing periodic boundary conditions along the $z$ direction, we introduce a perpendicular wave vector $k_z$, in addition to the small in-plane wave vector $\mathbf{q}$ measured from the $K$ point. The bulk Hamiltonian of the minimal model then reduces to a $2 \times 2$ matrix, 
\be 
H^{\rm{PBC}}(\mathbf{q},k_z) = \begin{pmatrix} 	0 &  v \pi^\dagger +\gamma_1 e^{i k_z d}\\   v \pi +\gamma_1 e^{-i k_z d} & 0 \end{pmatrix}, 
\ee 
where $\pi= \xi \hbar q_x+i  \hbar  q_y$ is measured with respect to the $K$ (for $\xi=1$) or $K'$ (for $\xi=-1$) point in the Brillouin zone. This dispersion has the nodal spiral~\cite{Heikkil__2011} in the 3D Brillouin zone at
\be
\label{eq:nodal}
\hbar v(\xi q_x-i q_y)=-\gamma_1 e^{i k_z d},
\ee
which corresponds to a circle of radius $q_0$ when projected to the $xy$ momentum plane, see Fig.~\ref{fig:helix}. 
Since the magnetic field points along $y$, we consider constant-energy contours at fixed $q_y$. Because the Fermi surface is concentrated around the Dirac spiral, we focus on low-energy solutions with $|q_y|<q_0$. For a given $q_y$, this fixes two possible values of $(q_x,k_z)$,
\bea
q_x^{({\rm{sp}})}&=&q_0 \tau \sqrt{1-(q_y/q_0)^2},~~~(\tau=\pm 1),\nonumber \\
k_z^{({\rm{sp}})}&=&\frac{1}{d} {\rm{arg}} \frac{- \xi q_x^{({\rm{sp}})} + i q_y}{q_0}.
\eea
The two values of $\tau$ correspond to the two branches of Landau levels observed in the spectrum. Together with the valley degree of freedom, this yields four points in momentum space where the nodal spiral intersects the fixed $k_y$ plane. Near each of these four points $(\xi,\tau)=(\pm,\pm)$, we expand the Hamiltonian to linear order in deviations from the Dirac spiral,
\bea
\xi&=&\pm,~~~q_y = {\rm{fixed}},~~~|q_y|<q_0, \nonumber \\
q_x&=&q_x^{({\rm{sp}})}+ \kappa_x  , \nonumber \\
k_z&=& k_z^{({\rm{sp}})} + \kappa_z,
\eea
and obtain 
\bea
H^{{\rm{PBC}}} \cong [v \xi \hbar \kappa_x - \gamma_1 \frac{q_y}{q_0} \kappa_z d]\sigma_x+[\gamma_1 \frac{\xi q_x^{({\rm{sp}})}  }{q_0}\kappa_z d]\sigma_y.
\eea
In standard notation, this is an anisotropic Dirac model
 \be 
 \label{eq:Pi}
 H^{{\rm{PBC}}}=  v  \Pi_x \sigma_x+ v_z  \Pi_z \sigma_y,
 \ee
 where $v_z=\frac{2}{\sqrt{3}}\frac{d}{a} \frac{\gamma_1}{\gamma_0} v \sqrt{1-\left( \frac{q_y}{q_0}\right)^2}$  and
 \bea
 \label{eq:Pix}
 \Pi_x=\hbar \left[\xi \left(\kappa_x + \frac{\hat{z}}{\ell_B^2}\right)-q_y d \kappa_z \right] ,~~~
 \Pi_z=\hbar \xi \tau \kappa_z,
 \eea
 satisfying $[\Pi_x,\Pi_z]=i \tau  \hbar e B$.

\subsection{Explicit solution}
The Hamiltonian Eq.~(\ref{eq:Pi}) can be solved exactly. We absorb the velocity-anisotropy into the momentum operators $\tilde{\Pi}_x= \Pi_x \sqrt{v/v_z}$, $\tilde{\Pi}_z= \Pi_z \sqrt{v_z/v}$ to obtain $H^{{\rm{PBC}}}=\sqrt{v v_z} (  \tilde{\Pi}_x \sigma_x +\tilde{\Pi}_z \sigma_y )$. Squaring, $(H^{{\rm{PBC}}})^2=v v_z \left(  \tilde{\Pi}_x^2 + \tilde{\Pi}_z^2  +i [\tilde{\Pi}_x,\tilde{\Pi}_z] \sigma_z \right)= v v_z \hbar e  B [2 a^\dagger a+1- \tau   \sigma_z]$ where we defined ladder operators
\be
a= \frac{\tilde{\Pi}_x+i \tau \tilde{\Pi}_z}{\sqrt{2\hbar e  B}},~~~a^\dagger= \frac{\tilde{\Pi}_x-i \tau \tilde{\Pi}_z}{\sqrt{2\hbar e  B}},
\ee
satisfying $[a,a^\dagger]=1$. We see that the spectrum is $E_n ={\rm{sgn}}(n)\sqrt{2 vv_z \hbar eB |n|}$, with $n=0,\pm 1, \pm 2,\dots$. It has not only a spin degeneracy and a valley degeneracy, but also the additional $\tau = \pm 1$ degeneracy resulting from the two intersections of the spiral with a constant $q_y$ plane.
In units of $\gamma_1$ as in the plots, 
\be
\label{eq:spec}
\left(\frac{E_n}{\gamma_1}\right)^2=2 \alpha |n|   \sqrt{1-\left(\frac{q_y}{q_0}\right)^2}.
\ee
We can see that the zero-th LL which is a zero eigenvector of the operator $2 a^\dagger a+1- \tau   \sigma_z$, has $\tau \sigma_z=1$. Namely it is localized in the $A$ sublattice for $\tau=1$  and vice versa. Indeed, as the tilted cylinder in Fig.~\ref{fig:tilted}(d) shows, the LL at $\tau=1$  connects to the top surface, for which the $B=0$ drumhead states are localized in the $A$-sublattice, and vice versa.

From Eq.~(\ref{eq:Pix}) it is apparent that for a bulk Landau eigenstate within the linearized theory the guiding centers  change along $z$ as
\be
\frac{d\langle z \rangle }{d \delta \kappa_x}= - \ell_B^2.
\ee
%where $\delta q_x = q_x - q_x^{(\rm sp)}$.

\section{zero energy states from SSH model}
\label{eq:SM_3_SSH}
Here, we follow Ref.~\cite{Tymczyszyn_2023} and map the minimal model of RMG onto an SSH chain. We then use this mapping to obtain the structure of the zero-energy states in $(q_x,q_y,z)$ space.

For zero magnetic field, the spectrum of the 3D RMG in the minimal model can be related to that of the 1D SSH model by treating both components of the in-plane wave vector $\mathbf{q}$ as parameters. In particular, the alternating hopping amplitudes $t$ and $w$ of the SSH model correspond to $\hbar v|\mathbf{q}|$ and $\gamma_1$, respectively, in RMG.

Ref.~\cite{Tymczyszyn_2023} generalized this correspondence to the presence of an in-plane magnetic field, resulting in an effective one-dimensional tight-binding model with spatially varying hopping amplitudes.
Starting from the Hamiltonian Eq.~(\ref{eq:2prime}) for $N$ layers of RMG, we gauge away the phase of the intralayer hopping term on layer $\ell$, leaving its magnitude
\be
v |\pi_\ell(q_x,q_y,z_n)|=\sqrt{(\hbar v q_x+e v z_\ell B_y)^2+(\hbar v q_y)^2} \equiv t_\ell,
\ee
with $B_x=0$. Treating $q_x$ and $q_y$ as parameters, the Hamiltonian in Eq.~(\ref{eq:2prime}) becomes an effective one-dimensional tight-binding model, which generalizes the SSH chain to $2N$ orbitals with hopping parameters ${ t_1,w,t_2,w,\dots, t_N }$, where $w=\gamma_1$. This model has a uniform intercell hopping $w$ and spatially modulated intracell hoppings $t_n$ on each unit cell.
 %The modulation is inversion symmetric about the center of the system, increasing in magnitude from $t_{N/2} \sim \hbar v |q|$ at the center to $t_{1,N} =\sqrt{(\hbar v q_y)^2+(\hbar v q_x+ev\frac{N}{2}dB)^2}$ at the edge.

Edge states of the SSH chain appear in the topological phase when $w>t_{1},t_{N}$, corresponding to states localized near the boundaries at $z=\pm d (N-1)/2$. Solitons of the SSH chain occur at positions $z_\ell$ and momenta $(q_x,q_y)$ for which $t_\ell=w$, namely
\be
\label{eq:z_0}
(\hbar v q_x+e v z_\ell B_y)^2+(\hbar v q_y)^2=\gamma_1^2.
\ee
In terms of the radius of the drumhead, $q_0$, the above condition can be rewritten as 
\be
\left(\frac{q_x}{q_0}+\alpha \left(\frac{N+1}{2}-\ell\right)\right)^2+\left(\frac{q_y}{q_0}\right)^2=1.
\ee
This describes a family of circles in momentum space, centered at $q_x/q_0=\alpha \left(\ell-(N+1)/2\right)$ with radius $q_0$, as shown in Fig.~\ref{fig:tilted}(d) in the main text. The surface formed by these tilted circles encodes the expectation value of the $z$ coordinate of zero-energy states as a function of $(q_x,q_y)$.

The SSH solitons coincide with the zero-th Landau levels found in our semiclassical solution, whose guiding centers move continuously from the bottom layer to the top layer as a function of the momentum $q_x$ perpendicular to the magnetic field. The states live on the lateral surface of the tilted cylinder in Fig.~\ref{fig:tilted}(d). The top and bottom bases of the tilted cylinder correspond to the  drumhead states at the top and bottom surfaces.

%Thus, as a function of $q_x$, the zero-energy state continuously changes its layer character. This state is doubly degenerate (in addition to spin and valley degeneracy), corresponding to the two possible orientations in which one can traverse the tilted cylinder from one surface to the other.

\section {Two-terminal conductance at fixed carrier density}
\label{sec:DC_conductivity_LB}
Here, we derive Eq.~(\ref{eq:G}) of the main text. This equation expresses the conductance in terms of the number of right-moving states at the chemical potential for each value of $q_y$.

We take the sample to be translationally invariant along \(y\), with
periodic boundary conditions over a width \(L_y\). The transverse momenta are
then quantized in steps \(\Delta q_y=2\pi/L_y\). For fixed valley \(\xi\),
spin \(s\), and transverse momentum \(q_y\), the slab is therefore an
effective one-dimensional conductor along $x$ with dispersions
\(E_{n\xi s}(q_x,q_y;B)\). Its Fermi points are the solutions of
\begin{equation}
E_{n\xi s}(q_x,q_y;B)=\mu .
\label{eq:Fermi_points}
\end{equation}
A Fermi point is right-moving when
\begin{equation}
v_{x,n\xi s}
=\frac{1}{\hbar}
\frac{\partial E_{n\xi s}}{\partial q_x}>0.
\end{equation}
We denote the number of such crossings at fixed \(q_y\) by
\begin{equation}
N_{R,\xi s}(\mu,q_y;B)
=
\sum_n\sum_{q_x^*}
\Theta\!\left[
\frac{\partial E_{n\xi s}}{\partial q_x}
(q_x^*,q_y;B)
\right],
\qquad
E_{n\xi s}(q_x^*,q_y;B)=\mu ,
\label{eq:NR_definition}
\end{equation}
where the sum is over simple Fermi crossings,
\(\partial_{q_x}E_{n\xi s}(q_x^*,q_y;B)\neq0\).

To see why each right-moving crossing contributes one conductance quantum, make a small change \(\delta\mu_S\) in the chemical potential of the source. The resulting current carried by the right-moving states is
\begin{align}
\delta I_{\xi s}
&=
-eL_y\delta\mu_S
\int\frac{dq_y}{2\pi}
\sum_n\int\frac{dq_x}{2\pi}\,
v_{x,n\xi s}\,
\Theta(v_{x,n\xi s})\,
\delta\!\left(\mu-E_{n\xi s}\right)
\nonumber\\
&=
-eL_y\delta\mu_S
\int\frac{dq_y}{2\pi}
\sum_{n,q_x^*}
\frac{1}{2\pi\hbar}
\Theta\!\left(v_{x,n\xi s}(q_x^*,q_y)\right)
\nonumber\\
&=
-\frac{e}{h}L_y\delta\mu_S
\int\frac{dq_y}{2\pi}\,
N_{R,\xi s}(\mu,q_y;B).
\label{eq:current_mode_counting}
\end{align}
In the second line we used
\[
\delta\!\left(\mu-E_{n\xi s}(q_x,q_y)\right)
=
\sum_{q_x^*}
\frac{\delta(q_x-q_x^*)}
{\left|\partial E_{n\xi s}/\partial q_x\right|_{q_x^*}},
\]
so that the group velocity cancels the one-dimensional density of states. The drain gives the corresponding contribution from left-moving states. For the complete slab spectrum, the occupied intervals at fixed \((\xi,s,q_y)\) have paired left- and right-moving boundaries, so \(N_{L,\xi s}=N_{R,\xi s}\).
Hence to linear order in the applied bias
\begin{equation}
I = -\frac{e}{h}L_y \sum_{\xi,s}
\int\frac{dq_y}{2\pi}\, N_{R,\xi s}(\mu,q_y;B)
\left(\mu_S-\mu_D\right).
\end{equation}
Using \(\mu_S-\mu_D=-eV_{\rm SD}\), the zero-temperature ballistic conductance is
therefore given by Eq.~\eqref{eq:G} which assumes ideal contacts and unit transmission.

\begin{figure*}[!h]
  \centering
\includegraphics[width=\textwidth]{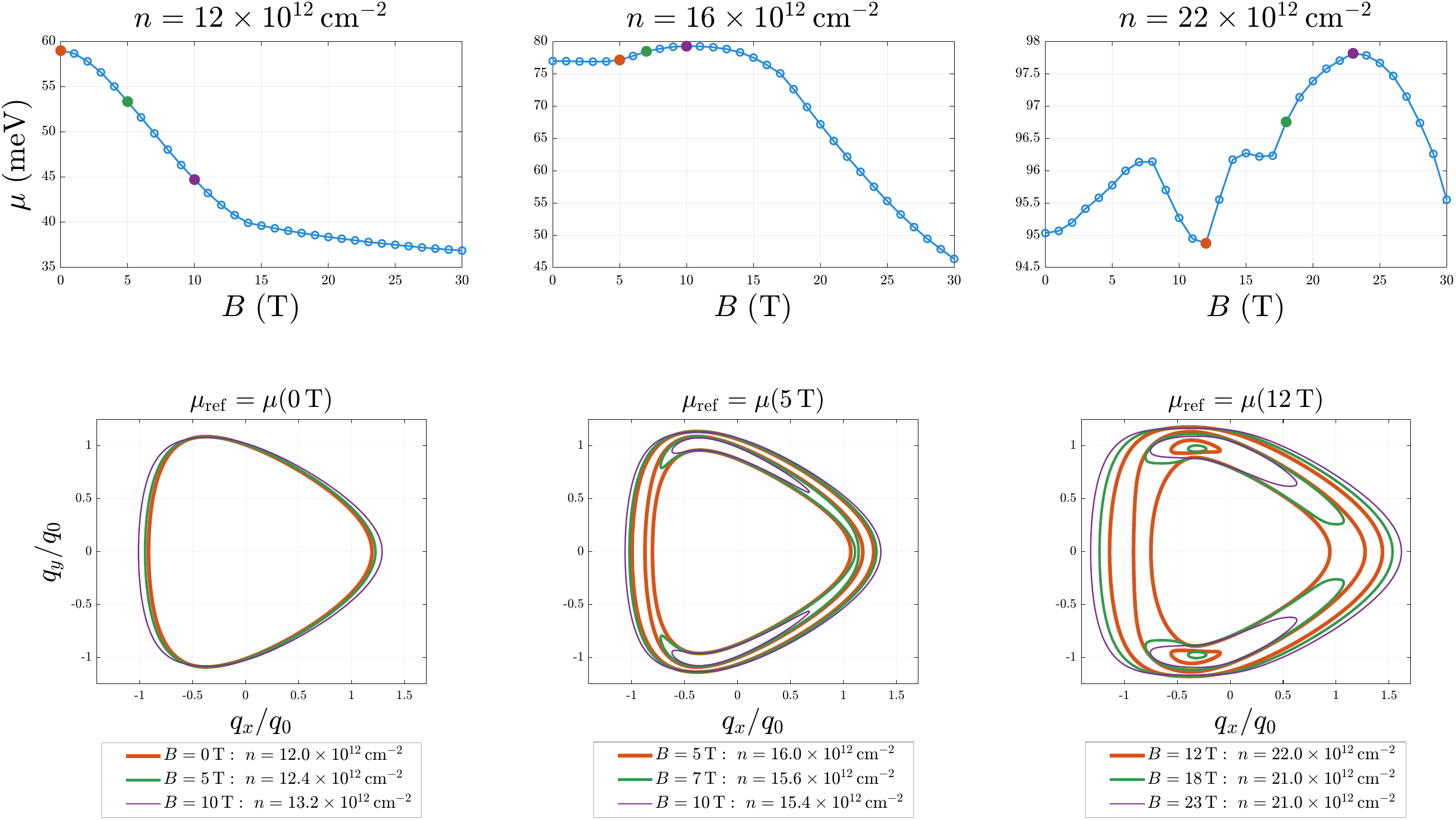}
  \caption{Chemical potential at fixed carrier density in an in-plane magnetic field for $N=51$. Top: \(\mu(B)\) for carrier densities \(n=(12,16,22)\times10^{12}\,\mathrm{cm}^{-2}\), from left to right, with Zeeman coupling neglected. Bottom: Fermi contours in valley \(\xi=+1\) at the fields marked by matching colors above. Within each panel, all contours are evaluated at the same reference chemical potential, \(\mu_{\mathrm{ref}}=\mu(B_{\mathrm{ref}})\), with \(B_{\mathrm{ref}}=0,5,12\,\mathrm T\), respectively. The legends give the corresponding carrier densities. Left: increasing \(B\) expands the occupied momentum-space area at fixed \(\mu_{\mathrm{ref}}\), consistent with the growing number of low-energy states associated with the emerging zeroth Landau level. Maintaining a fixed density therefore requires a decrease in \(\mu\). Center: the inner nested contours reconnect into separate pockets that shrink with increasing field. Their loss of occupied area outweighs the expansion of the outer contour, reducing the density at fixed \(\mu_{\mathrm{ref}}\) and requiring \(\mu\) to increase. Right: the small pockets present at \(12\,\mathrm T\) shrink and eventually disappear, while the larger nested contours reconnect into pockets that subsequently shrink. The resulting net loss of occupied area likewise drives an increase in \(\mu\) at fixed density.}
  \label{fig: mu_B N51}
\end{figure*}

When a field sweep is performed at fixed carrier density $n_0$, the chemical potential is determined by $n\lp \mu(B), B \rp = n_0$ where
\begin{equation}
n(\mu,B) = \sum_{\xi,s}\int\frac{d^2q}{(2\pi)^2}
\left[ \sum_{a=1}^{2N}f\bigl(E_{a\xi s}(\bm q,B)-\mu\bigr)-N \right].
\label{eq:density_full_spectrum}
\end{equation}
Here \(n\) is the excess sheet density relative to half filling, and the subtraction of \(N\) removes the half-filled reference occupation in each spin–valley sector. At zero temperature, \(f(E-\mu)=\Theta(\mu-E)\).
Examples are shown in Fig.~\ref{fig: mu_B N51}. We emphasize the non-monotonic behavior of $\mu(B)$. To explain the non-monotonicity we use
\begin{equation}
\left(\frac{\partial\mu}{\partial B}\right)_n
= -\frac{(\partial n/\partial B)_\mu}{(\partial n/\partial\mu)_B}. 
\label{eq:dmu_dB_fixed_n}
\end{equation}
 
The denominator is the thermodynamic density of states and is positive whenever the system is compressible. The numerator, however, has no fixed sign because the magnetic field reconstructs the spectrum. At zero temperature,
\begin{equation}
\left(\frac{\partial n}{\partial B}\right)_{\mu}
=
-\sum_{\xi,s,n}\int\frac{d^2q}{(2\pi)^2}\,
\delta\!\left[\mu-E_{n\xi s}(\bm q;B)\right]
\frac{\partial E_{n\xi s}(\bm{q}, B)}{\partial B}.
\end{equation}
It is therefore determined by a Fermi-surface average of the field-induced energy shifts, which may have either sign. If the field shifts states downward through the fixed chemical potential or enlarges the occupied Fermi-sea area, then \((\partial n/\partial B)_\mu>0\) and \(\mu(B)\) decreases at fixed density. Conversely, upward spectral shifts or shrinking Fermi pockets give
\((\partial n/\partial B)_\mu<0\) and an increasing \(\mu(B)\). Different bands and regions of the Fermi contours can contribute with opposite signs, and their changing balance produces the nonmonotonic dependence of \(\mu\) on \(B\).

\FloatBarrier

\section{Layer-resolved current operator}
\label{se:layer_resolved_I}
Here, we present calculations of the layer-resolved current and show that, while weak-field finite slabs can carry a current peaked in the interior, field-induced redistribution can produce a current distribution concentrated near the surfaces.

Let $P_\ell$ denote the projector onto layer $\ell$. In the basis $(A_1,B_1,A_2,B_2,\ldots,A_N,B_N)$, it is given by $P_\ell =
|A_\ell\rangle\langle A_\ell| + |B_\ell\rangle\langle B_\ell|$. The projectors satisfy $  \sum_{\ell=1}^{N}P_\ell=\mathbbm{1}$. The total single-particle current operator along $x$ is $\hat J_x=-e\hat v_x$,
where $ \hat v_x = \frac{1}{\hbar} \frac{\partial H(\mathbf q)}{\partial q_x}$. We define the current associated with layer $\ell$ as
\begin{equation}
\hat J_x^{(\ell)} = -\frac{e}{2} \left\{ P_\ell,\hat v_x \right\}.
\label{eq:layer_current_operator}
\end{equation}
The anticommutator ensures that the layer-resolved current operator is
Hermitian even when $P_\ell$ and $\hat v_x$ do not commute.
With this definition the layer currents add exactly to the total current
\begin{align}
\sum_{\ell=1}^{N}\hat J_x^{(\ell)} = -\frac{e}{2} \sum_{\ell}
\left( P_\ell\hat v_x+\hat v_xP_\ell \right) =
-\frac{e}{2} \left(\hat v_x+\hat v_x\right)  = -e\hat v_x = \hat J_x.
\end{align}

For the minimal rhombohedral graphene model considered here, the
$q_x$ dependence resides in the intralayer Dirac Hamiltonians, whereas the
dominant interlayer hopping $\gamma_1$ is independent of $q_x$. Consequently,
$\hat v_x$ is diagonal in layer space and $[P_\ell,\hat v_x]=0$. Eq.~(\ref{eq:layer_current_operator}) therefore simplifies to
$\hat J_x^{(\ell)} = -eP_\ell\hat v_xP_\ell$.

For an eigenstate $|n\mathbf q\rangle$, we define
\begin{equation}
J_{x,n\mathbf q}^{(\ell)} = \langle n\mathbf q|\hat J_x^{(\ell)}|n\mathbf q\rangle.
\end{equation}
Summing over layers gives
$\sum_\ell J_{x,n\mathbf q}^{(\ell)} = -e\langle n\mathbf q|\hat v_x|n\mathbf q\rangle = -\frac{e}{\hbar} \frac{\partial E_n(\mathbf q)}{\partial q_x} = -e v_{x,n\mathbf q}$, where the second equality follows from the Hellmann--Feynman theorem. 

In the Landauer--Imry--B\"uttiker geometry, the two reservoirs are attached to
opposite $x$ faces and have electrochemical potentials $    \mu_L-\mu_R=eV$. For an ideal ballistic system, right-moving states are populated by the
left reservoir and left-moving states by the right reservoir,
\begin{equation}
f_{n\mathbf q}^{\rm LB} =
\begin{cases}
    f_L(E_{n\mathbf q}), & v_{x,n\mathbf q}>0, \\[1mm]
    f_R(E_{n\mathbf q}), & v_{x,n\mathbf q}<0.
\end{cases}
\end{equation}
The layer-resolved current density is therefore
\begin{equation}
j_{x,\rm tot}^{\rm LB}(z_\ell) = \frac{1}{A}    \sum_{n,\mathbf q} f_{n\mathbf q}^{\rm LB}
J_{x,n\mathbf q}^{(\ell)},
\label{eq:LB_layer_current}
\end{equation}
where $A=L_xL_y$ is the in-plane area. For a symmetrically applied small bias, $\mu_L=\mu+\frac{eV}{2},~ \mu_R=\mu-\frac{eV}{2}$, the linear-response change in occupation is
\begin{equation}
\delta f_{n\mathbf q}^{\rm LB} =  -\frac{\partial f}{\partial E}
\frac{eV}{2} \operatorname{sgn} \left(v_{x,n\mathbf q}\right).
\end{equation}
At zero temperature $-\frac{\partial f}{\partial E} = \delta(E-\mu)$, and the current induced to linear order in the applied bias is
\begin{equation}
j_x^{\rm LB}(z_\ell) = \frac{eV}{2A} \sum_{n,\mathbf q}
\delta(E_{n\mathbf q}-\mu) \operatorname{sgn} \left(v_{x,n\mathbf q}\right)
J_{x,n\mathbf q}^{(\ell)}.
\label{eq:LB_layer_current_zeroT}
\end{equation}
Thus the Landauer--Imry--B\"uttiker transport-current profile is determined by
the propagating states at the Fermi energy. 

We now examine how the Landauer- Imry-B\"uttiker transport current is distributed
across the thickness of the slab. To compare slabs with different numbers of
layers, we define the normalized layer-resolved profile
\begin{equation}
\widetilde{j}^{\rm LB}_x(z_\ell) = \frac{j^{\rm LB}_x(z_\ell)}
{\displaystyle\sum_{\ell'=1}^{N} j^{\rm LB}_x(z_{\ell'})}.
\label{eq:normalized_LB_layer_current}
\end{equation}

\begin{figure*}[!h]
  \centering
\includegraphics[width=\textwidth]{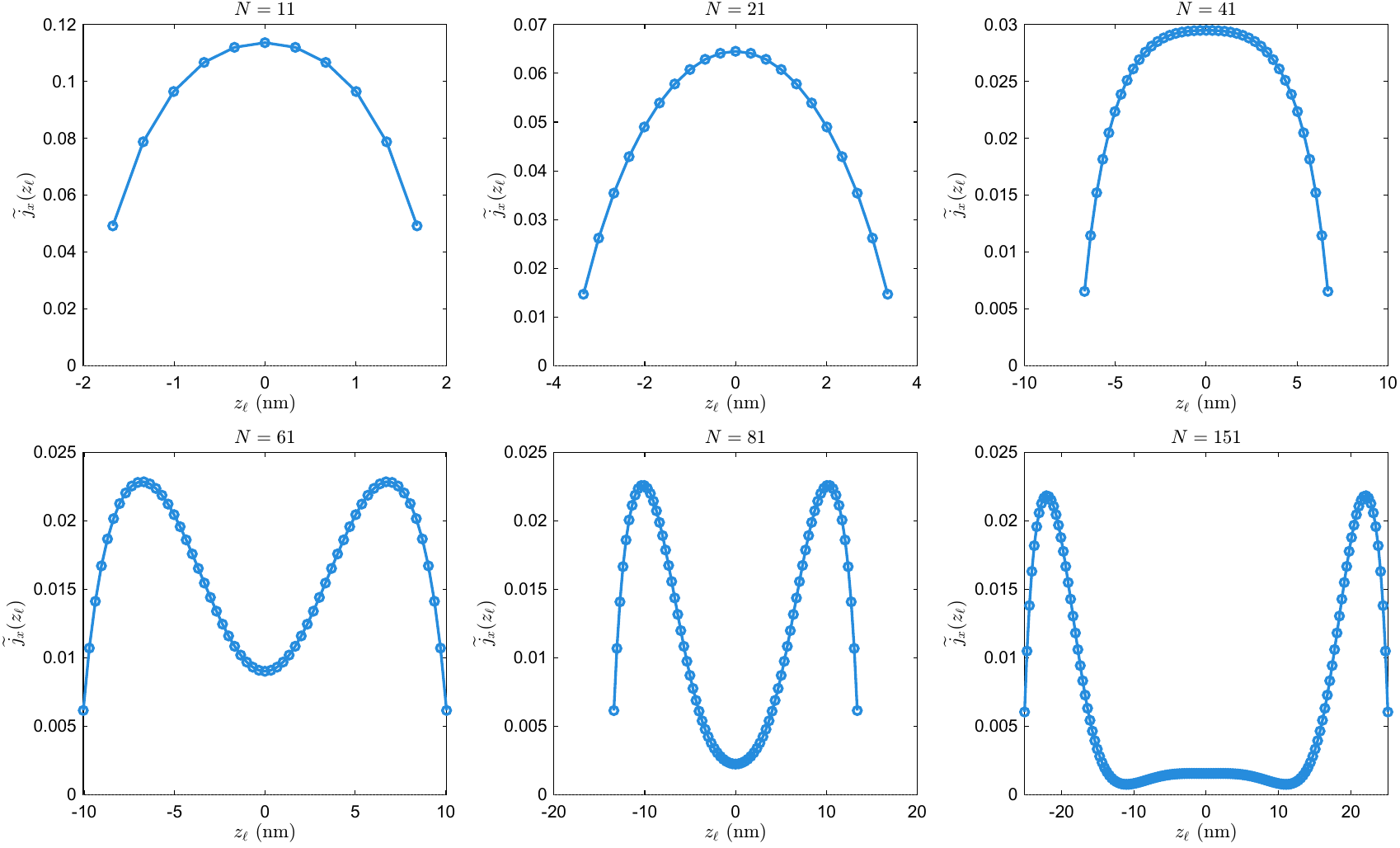}
  \caption{Normalized layer-resolved transport-current profiles for the minimal model at fixed magnetic field \(B=10\,\mathrm T\) and chemical potential \(\mu=10\,\mathrm{meV}\), for the slab thicknesses indicated. Each profile is normalized such that \(\sum_\ell\widetilde j_x(z_\ell)=1\). For thin slabs, the current is broadly distributed and maximal near the center. As the thickness increases, the central maximum broadens and splits into two subsurface maxima near opposite surfaces, while the current in the interior is suppressed. The current-carrying regions retain a finite penetration depth into the slab.}
  \label{fig:LB_layer_current}
\end{figure*}

Fig.~\ref{fig:LB_layer_current} shows
$\widetilde{j}^{\rm LB}_x(z_\ell)$ for several slab thicknesses for $B=10{\rm{T}}$. The current
distribution undergoes a pronounced crossover as $N$ increases. For the
thinnest slabs, such as $N=11$ and $N=21$, the current is broadly distributed
across the slab and is maximal near its center. 
At intermediate thickness, the profile first broadens and then changes its
shape. Around \(N=41\), it remains approximately center-peaked, whereas by
\(N=61\) the central maximum has split into two off-center maxima. Upon
increasing \(N\) further, these maxima move toward the opposite surfaces and
the current carried through the central region is progressively suppressed.
For \(N=81\), and \(151\), most of the transport current is
concentrated in two well-separated regions associated with the top and bottom
surfaces. Nevertheless the two peaks in current are not restricted to the
outermost layers. 

%\eran{We note that from the condition $\Delta_N=\Delta_B$ we obtain the typical crossover number of layers $N^*=11.5$} which is lower than the actual crossover towards edge transport.

\FloatBarrier

\section{Surface Hall voltage} 
\label{se:surface_voltage}
We derive the equations describing the 4-terminal response.
A weak voltage probe samples states incident from the source and
drain without appreciably perturbing their occupations. For equal
incoherent coupling to the two orbitals of the outermost layers,
we define
\begin{align}
P_T&=|A_1\rangle\langle A_1|+|B_1\rangle\langle B_1|,\\
P_B&=|A_N\rangle\langle A_N|+|B_N\rangle\langle B_N|.
\end{align}
The weight of a slab eigenstate on the contacted orbitals is
\[
w_{pn\xi s}(\bm q) = \langle n\xi s,\bm q|P_p|n\xi s,\bm q\rangle, \qquad p=T,B.
\]
At zero temperature, the directional local densities of states are
\begin{equation}
\nu_p^\pm = \sum_{\xi,s,n} \int\frac{d^2q}{(2\pi)^2}\, w_{pn\xi s}(\bm q)\,
\Theta(\pm v_{x,n\xi s})\, \delta\!\left[\mu-E_{n\xi s}(\bm q)\right].
\label{eq:probe_DOS_2D}
\end{equation}
At fixed \(q_y\), the delta function can be expanded over simple
Fermi crossings,
\begin{equation}
\delta\!\left[\mu-E_{n\xi s}(q_x,q_y)\right]
=
\sum_{a\in n}
\frac{\delta(q_x-q_{x,a})}{\hbar|v_{x,a}|}.
\end{equation}
Integration over \(q_x\) yields
\begin{equation}
\label{eq:probe1}
\nu_p^\pm
=
\sum_{\xi,s}\int\frac{dq_y}{(2\pi)^2}
\sum_{a:\,\pm v_{x,a}>0}
\frac{w_{pa}}{\hbar|v_{x,a}|}.
\end{equation}

Let the probe be at chemical potential $\mu_p$. In the ballistic limit, right movers are populated by the source and left movers by the drain. To leading order in the probe coupling and in the applied bias, the tunneling current is
\begin{equation}
I_p=C_p\left[ \nu_p^+(\mu_p-\mu_S) + \nu_p^-(\mu_p-\mu_D) \right],
\end{equation}
where \(C_p>0\) contains a common tunneling prefactor. The floating-probe condition \(I_p=0\) gives
\be
\mu_p=\frac{\nu_p^+ \mu_S+\nu_p^- \mu_D}{\nu_p^+ +\nu_p^-};
\ee
the chemical potential of the probe is the average of the source and drain chemical potentials weighted by the directional densities.
Writing \(\mu_{S,D}=\bar\mu\pm eV_{\rm SD}/2\), we obtain
\begin{equation}
\mu_p=\bar\mu+\frac{eV_{\rm SD}}{2}\eta_p, \qquad  \text{with} \qquad
\eta_p = \frac{\nu_p^+-\nu_p^-}{\nu_p^++\nu_p^-},
\end{equation}
where \(-1\leq\eta_p\leq1\) describes the difference between the coupling of probe $p$ to right movers and left movers. Subtracting the top and bottom probe potentials and using
\(I=GV_{\rm SD}\) and \(eV_H=\mu_T-\mu_B\) gives
\begin{equation}
\label{eq:Hall1}
\frac{V_H}{V_{\rm SD}} = \frac{\eta_T-\eta_B}{2},
\qquad
R_H = \frac{1}{G}\frac{\eta_T-\eta_B}{2}.
\end{equation}
When the top and bottom probes couple exclusively to oppositely propagating states, \(\eta_T=-\eta_B=\pm1\), then \(|V_H/V_{\mathrm{SD}}|=1\) and \(|R_H|=1/G\).

\section{Bulk Hall conductivity}
\label{sec:slab_conductivity}
We provide a detailed derivation of Eq.~(\ref{eq:sigma_xz}) in the main text from the Kubo formula for the transverse conductivity, in terms of the exact eigenstates of a finite rhombohedral slab.

For each valley $\xi$, the exact $2N\times 2N$ slab Hamiltonian satisfies
\begin{equation}
H_N^{(\xi)}(\bm q)
\ket{u_{n\xi}(\bm q)}
=
E_{n\xi}(\bm q)
\ket{u_{n\xi}(\bm q)},
\qquad
\bm q=(q_x,q_y),
\label{eq:slab_eigenstates_conductivity}
\end{equation}
where $n=1,\ldots,2N$. The eigenstates are normalized according to
$\braket{u_{n\xi}(\bm q)|u_{m\xi}(\bm q)} = \delta_{nm}$. The position operator along the stacking direction is
\begin{equation}
\hat Z = \sum_{\ell=1}^{N} z_\ell
\left(\ket{\ell A}\bra{\ell A} + \ket{\ell B}\bra{\ell B} \right).
\label{eq:Z_operator}
\end{equation}
The velocity operators parallel and perpendicular to the layers are,
respectively,
\bea
\hat v_x(\bm q) &=& \frac{1}{\hbar} \frac{\partial H_N(\bm q)}{\partial q_x}, \label{eq:vx_slab} \\
\hat v_z(\bm q) &=& \frac{i}{\hbar} \left[ H_N(\bm q),\hat Z \right].
\label{eq:vz_slab}
\eea
We define the corresponding matrix elements
\begin{align}
v^i_{nm,\xi}(\bm q)
&=
\bra{u_{n\xi}(\bm q)}
\hat v_i(\bm q)
\ket{u_{m\xi}(\bm q)},
\\
Z_{nm,\xi}(\bm q)
&=
\bra{u_{n\xi}(\bm q)}
\hat Z
\ket{u_{m\xi}(\bm q)}.
\end{align}
Using Eq.~\eqref{eq:slab_eigenstates_conductivity}, the matrix elements of the perpendicular velocity satisfy
\begin{align}
v^z_{nm,\xi}(\bm q)
=
\frac{i}{\hbar}
\bra{u_{n\xi}}
\left[
H_N,\hat Z
\right]
\ket{u_{m\xi}}
=
\frac{i}{\hbar}
\left[
E_{n\xi}(\bm q)-E_{m\xi}(\bm q)
\right]
Z_{nm,\xi}(\bm q).
\label{eq:vz_Z_relation}
\end{align}

\subsection{Kubo formula}

The sheet conductivity of the slab is
\begin{align}
\Sigma_{ij}(\omega) = -i e^2\hbar
\sum_{\xi,s} \int\frac{d^2q}{(2\pi)^2}
\sum_{n\neq m}  \frac{f_{n\xi}(\bm q)-f_{m\xi}(\bm q)}{
E_{n\xi}(\bm q)-E_{m\xi}(\bm q)} \times \frac{
v^i_{nm,\xi}(\bm q) v^j_{mn,\xi}(\bm q)}{E_{n\xi}(\bm q)-E_{m\xi}(\bm q) +\hbar\omega+i\Gamma}.
\label{eq:slab_kubo_general}
\end{align}
Here $\Gamma>0$ is a phenomenological broadening.
We take the magnetic field along the $y$ direction, and 
an electric field $\bm E=E_z\hat{\bm z}$ applied perpendicular to the stack. The off diagonal sheet conductivity is then
\be
\Sigma_{xz}(\omega,B) = -i e^2\hbar
\sum_{\xi,s} \int\frac{d^2q}{(2\pi)^2} \sum_{n\neq m}
\frac{f_{n\xi}-f_{m\xi}}{E_{n\xi}-E_{m\xi}}
\frac{v^x_{nm,\xi}v^z_{mn,\xi}}{E_{n\xi}-E_{m\xi}+\hbar\omega+i\Gamma}.
\label{eq:sigma_xz_velocity}
\ee
Using Eq.~(\ref{eq:vz_Z_relation}), the energy difference in the first denominator cancels, yielding
\begin{equation}
\Sigma_{xz}(\omega, B) =e^2 \sum_{\xi,s}
\int\frac{d^2q}{(2\pi)^2} \sum_{n\neq m}
\left(f_{n\xi}-f_{m\xi}\right)
\frac{ Z_{nm,\xi} v^x_{mn,\xi}}{-\lp E_{n\xi}-E_{m\xi} \rp +\hbar\omega+i\Gamma
}. 
\label{eq:sigma_xz_Z}
\end{equation}
We now take the limit of DC Hall conductivity. Setting $\omega=0$ as well as $\Gamma=0$, and using 
\be
v^x_{nm,\xi} = \lp v^x_{mn,\xi} \rp^\star  \qquad
Z_{mn,\xi} = \lp Z_{nm,\xi} \rp^\star
\ee
we obtain 
\begin{equation}
\Sigma_{xz}^{\rm DC} = -2e^2 \sum_{\xi,s}
\int\frac{d^2q}{(2\pi)^2} \sum_n f_{n\xi}
\sum_{m\neq n} \frac{\operatorname{Re}\left[
v^x_{nm,\xi}Z_{mn,\xi} \right] }{E_{n\xi}-E_{m\xi} }.
\label{eq:sigma_xz_dc_Z}
\end{equation}
The remaining energy denominator can be eliminated using the momentum
dependence of the exact eigenstates.  Differentiating $H_N(\bm q)\ket{u_{m\xi}(\bm q)}
=
E_{m\xi}(\bm q)\ket{u_{m\xi}(\bm q)}$ with respect to $q_x$ and projecting onto
$\bra{u_{n\xi}}$, with $m\neq n$, gives
\begin{equation}
\bra{u_{n\xi}}
\frac{\partial H_N}{\partial q_x}
\ket{u_{m\xi}}
=
\left(E_{m\xi}-E_{n\xi}\right)
\braket{u_{n\xi}|\partial_{q_x}u_{m\xi}}.
\label{eq:eigenstate_derivative}
\end{equation}
Since
\begin{equation}
v^x_{nm,\xi}
=
\frac{1}{\hbar}
\bra{u_{n\xi}}
\frac{\partial H_N}{\partial q_x}
\ket{u_{m\xi}},
\end{equation}
we have
\begin{equation}
\frac{v^x_{nm,\xi}}{E_{n\xi}-E_{m\xi}}
= -\frac{1}{\hbar} \braket{u_{n\xi}|\partial_{q_x}u_{m\xi}}.
\label{eq:vx_energy_relation}
\end{equation}
Therefore
\begin{equation}
\Sigma_{xz}^{\rm DC} = \frac{2e^2}{\hbar } \sum_{\xi,s}
\int\frac{d^2q}{(2\pi)^2} \sum_n f_{n\xi}
\sum_{m\neq n} \operatorname{Re} \left[ \braket{u_{n\xi}|\partial_{q_x}u_{m\xi}} Z_{mn,\xi} \right].
\label{sigma_xz2}
\end{equation}
Adding the $m=n$ term to the sum does not change its real part, since
$\braket{u_{n\xi}|\partial_{q_x}u_{n\xi}}$ is purely imaginary while
$Z_{nn,\xi}$ is real.  Using $\braket{u_{n\xi}|\partial_{q_x}u_{m\xi}}
=
-\braket{\partial_{q_x}u_{n\xi}|u_{m\xi}}$,  we find
\begin{align}
\sum_m
\operatorname{Re}\left[
\braket{u_{n\xi}|\partial_{q_x}u_{m\xi}}
Z_{mn,\xi}
\right]
&=
-\operatorname{Re}
\bra{\partial_{q_x}u_{n\xi}}
\hat Z
\ket{u_{n\xi}}.
\end{align}
Defining
\begin{equation}
z_{n\xi}(\bm q)
\equiv
Z_{nn,\xi}(\bm q)
=
\bra{u_{n\xi}(\bm q)}
\hat Z
\ket{u_{n\xi}(\bm q)},
\end{equation}
and noting that $\hat Z$ is independent of $q_x$, we have $\frac{\partial Z_{n\xi}}{\partial q_x}
=
2\operatorname{Re}
\bra{\partial_{q_x}u_{n\xi}}
\hat Z
\ket{u_{n\xi}}$.
 Therefore
\begin{equation}
\Sigma_{xz}^{\rm DC}
=
-\frac{e^2}{\hbar}
\sum_{\xi,s}
\int\frac{d^2q}{(2\pi)^2}
\sum_n
f_{n\xi}(\bm q)
\frac{\partial z_{n\xi}(\bm q)}{\partial q_x}.
\end{equation}
The three-dimensional conductivity is obtained by dividing the current by the full slab volume,  $\sigma_{ij}(\omega) = \frac{\Sigma_{ij}(\omega)}{L_z}$, where $L_z$ is the slab thickness. This gives Eq.~(\ref{eq:sigma_xz}).

At zero temperature $f_{n\xi}=\Theta(\mu-E_{n\xi})$. Define the q-dependent stacking position 
\begin{equation}
Z_{{\rm occ},\xi}(\bm q)=\sum_n f_{n\xi}(\bm q)\,z_{n\xi}(\bm q),
\label{eq:Zocc}
\end{equation}
so that $-eZ_{{\rm occ},\xi}$ is the dipole moment of the electrons at momentum $\bm q$. Differentiating,
\begin{equation}
\frac{\partial Z_{{\rm occ},\xi}}{\partial q_x}
=\sum_n f_{n\xi}\frac{\partial z_{n\xi}}{\partial q_x}
-\sum_n\delta(\mu-E_{n\xi})\frac{\partial E_{n\xi}}{\partial q_x}\,z_{n\xi}.
\end{equation}
At fixed $q_y$, let $q_x^{\rm L}$ and $q_x^{\rm R}$ lie beyond all Fermi crossings. Integrating, and using
$\delta(\mu-E_{n\xi})=\sum_a\delta(q_x-q_{x,a})/|\partial_{q_x}E_{n\xi}|$ for the simple crossings $q_{x,a}$ of band $n$,
\begin{equation}
\int_{q_x^{\rm L}}^{q_x^{\rm R}}dq_x\sum_n f_{n\xi}\frac{\partial z_{n\xi}}{\partial q_x}
=Z_{{\rm occ},\xi}\Big|_{q_x^{\rm L}}^{q_x^{\rm R}}
+\sum_a{\rm sgn}(v_{x,a})\,z_a ,
\label{eq:parts_SM}
\end{equation}
where $a$ runs over all Fermi crossings at fixed $q_y$ (all bands), $v_{x,a}=\hbar^{-1}\partial_{q_x}E_{n\xi}$ and $z_a=z_{n\xi}(q_{x,a},q_y)$ are the velocity and position of the crossing state. 

We now show that the first term in Eq.~\eqref{eq:parts_SM} vanishes. At $q_x^{\rm L,R}$ the chemical potential lies in a gap, so the occupied states are the $N$ lowest bands and $Z_{{\rm occ},\xi}=\sum_\ell z_\ell\,n_\ell$, with $n_\ell$ the occupied charge on layer $\ell$. In the minimal model the
Hamiltonian is chiral, $\Gamma H_N\Gamma=-H_N$ with $\Gamma=\pm1$ on the $A$ ($B$) sites, so $\Gamma\ket{u_{n\xi}}$ is an eigenstate at $-E_{n\xi}$ with the same weight on every site. $\Gamma$ thus maps the occupied (negative-energy) states onto the empty (positive-energy) ones without changing the site weights, so each site has
equal occupied and empty weight, which add up to one by completeness. Each site is therefore half filled, $n_\ell=1$ for all $\ell$, and $Z_{{\rm occ},\xi}=\sum_\ell z_\ell=0$ at both ends.

Physically, the filled valence band is neutral layer by layer and carries no dipole; a dipole appears only when states near $\mu$ are added or removed, which happens at the Fermi crossings. The remote hopping $\gamma_4$ $\lp \gamma_4/\gamma_0 \approx 0.04 \rp$ breaks the chiral symmetry, but only weakly, and we verified numerically that  $|Z_{{\rm occ},\xi}(q_x^{\rm L,R})|$ is negligible for all fields and $q_y$ considered. Therefore, with 
\begin{equation}
S_N(q_y) \equiv -\frac{1}{L_z}\sum_a{\rm sgn}(v_{x,a})\,z_a ,
\end{equation}
we obtain Eq.~(\ref{eq:sigma_xz}).

The same steps apply to a two-dimensional electron gas confined to $0<y<W$ in a perpendicular field, with $\hat Z\to\hat y$, $L_z\to W$ and a single conserved momentum $k_x$. For $\mu$ in a Landau gap the boundary term in Eq.~(\ref{eq:parts_SM}) vanishes trivially, since beyond the outermost crossing all levels lie above $\mu$, and one obtains $\sigma_{xy}=(e^2/h)\sum_{n\le\nu}[y_n(k_R)-y_n(k_L)]/W$, with $k_{L,R}$ the two Fermi crossings of level $n$. For a hard wall the crossing states lie $\sim\ell_B$ inside the edges, so $\sigma_{xy}=\nu e^2/h\,[1-O(\ell_B/W)]$, whereas an edge probe exchanges electrons with a single chirality, reads $\mu_S$ or $\mu_D$, and gives $R_H=h/\nu e^2$ exactly. 

For $|q_y|<q_0$ the bulk problem at fixed $q_y$ is the pair of anisotropic Dirac cones $\tau=\pm$ of Eq.~(\ref{eq:Pi}) in the common field $B_y$. The two cones have opposite chirality, like the $K$ and $K'$ valleys of graphene, and, as there, their Hall conductivities add rather than cancel. With the $n=0$ level filled each contributes $e^2/2h$ with a sign set by the
field~\cite{Zheng2002}, so the slice carries $C(q_y)=1$ per spin and valley for $\mu$ between the zeroth and first levels. For $|q_y|>q_0$ the plane misses the spiral, Eq.~(\ref{eq:nodal}), and $C=0$; the change is permitted by the closing of the Landau gap at $q_0$, Fig.~\ref{fig:3_SM}. Bulk--boundary correspondence~\cite{Hatsugai1993} applies once the slab
has a bulk, i.e. once Landau levels have formed inside it, $\Delta_B\gg\Delta_N$ ($L_z\gg\delta_z$), so that the modes on the two surfaces are decoupled. It then gives $N_R-N_L=\pm C(q_y)$ on the two surfaces, which fixes $N_R(q_y)=1$ in Eq.~(\ref{eq:G}) and $\eta_{t,b}=\mp1$ in Eq.~(\ref{eq:VH}). For $\Delta_B\lesssim\Delta_N$ the two surfaces hybridize, the spectrum consists of confinement subbands, and $N_R$ is not
fixed by $C(q_y)$; this is the low-field regime of Figs.~\ref{fig:results} and \ref{fig:Kubo}.

\section{Zero-energy states}
\label{sec:zero_modes}

Here we count the zero-energy states of the minimal model whose locus is the tilted cylinder of Fig.~\ref{fig:tilted}(d), and define the layer-resolved spectral function of Fig.~\ref{fig:tilted}(c). All counts are per spin and valley.

\paragraph{Cylinder wall.} At fixed $q_y$ with $|q_y|<q_0$, each of the two zeroth-Landau-level branches $\tau=\pm$ exists over the range of $q_x$ in which its guiding center lies inside the slab. Since $d\langle z\rangle/dq_x=-\ell_B^2$ (Sec.~\ref{eq:SM_2_semiclassical}), this range is $\Delta q_x=L_z/\ell_B^2=\alpha (N-1)q_0$, the same for all $|q_y|<q_0$. The number of wall states per unit area of the slab is therefore
\be
\int_{-q_0}^{q_0}\frac{dq_y}{2\pi}\;2\,\frac{\alpha Nq_0}{2\pi}=\frac{\alpha N q_0^2}{\pi^2},
\ee
and per unit volume, dividing by $L_z=(N-1)d$ and using $\alpha=d/\ell_B^2q_0$,
\be
\rho_{3D}=\frac{q_0}{\pi^2\ell_B^2}
=2\times\frac{1}{2\pi\ell_B^2}\times\frac{2q_0}{2\pi}:
\ee
the Landau degeneracy of each $xz$ plane, times the density $q_0/\pi$ of $q_y$ channels along the field, times two for $\tau=\pm$. It is proportional to $B$ and independent of $N$.

\paragraph{Bases.} Each drumhead disk has radius $q_0$ in momentum space, Sec.~\ref{eq:SM_3_SSH}, and contributes $\pi q_0^2/(2\pi)^2=q_0^2/4\pi$ states per unit area; the two bases together $q_0^2/2\pi$, independent of $B$. \\

\noindent
The number of surface to bulk zero-energy states is
\be
\frac{q_0^2/2\pi}{\alpha Nq_0^2/\pi^2}=\frac{\pi}{2\alpha N},
\ee
equal to $1.46$ for $N=51$ at $25$~T. In this geometric estimate, the number of bulk zeroth-Landau-level states exceeds the number of drumhead surface states when \(\alpha N>\pi/2\).

\paragraph{Spectral function.} Fig.~\ref{fig:tilted}(c) shows
\be
A(E,\bm q,z_\ell)=-\frac1\pi\,{\rm Im}\sum_{n=1}^{2N}\frac{\sum_{\alpha=A,B}|u^{\,n}_{\ell\alpha}(\bm q)|^2}{E-E_n(\bm q)+i\Gamma},
\ee
with $u^{\,n}_{\ell\alpha}$ the amplitude of eigenstate $n$ on site $\alpha$ of layer $\ell$. In a finite slab a Landau branch arriving at a surface meets the drumhead state of that surface near the edge of its disk; the two hybridize and split. The weight at $E=0$ therefore vanishes over a range $\sim\delta_z/\ell_B^2$ of $q_x$ at each end of the branch. The wall of the cylinder is the locus of the zero-energy states, not a surface they cover.

\end{document}